\documentclass[12pt]{article}
 \usepackage{amsmath,amssymb,graphicx,ulem,mathrsfs,caption,color}
 \usepackage{bm}
 \usepackage{graphicx}
 \usepackage{subfigure}
 \usepackage{epstopdf}
 \usepackage{float}
 \usepackage{colordvi,multicol}
\usepackage{color}
\usepackage{multirow}
  \usepackage{amsthm}
  \usepackage[margin=1in]{geometry}
 \allowdisplaybreaks
 
 \newtheorem{theorem}{Theorem}
\usepackage[symbol]{footmisc}

\newcommand{\uzero}       {\mbox{\boldmath$0$}}

\newcommand{\uB}       {\mbox{\boldmath$B$}}

\newcommand{\uf}       {\mbox{\boldmath$f$}}
\newcommand{\uG}       {\mbox{\boldmath$G$}}

\newcommand{\uI}       {\mbox{\boldmath$I$}}

\newcommand{\uJ}       {\mbox{\boldmath$J$}}

\newcommand{\uQ}       {\mbox{\boldmath$Q$}}

\newcommand{\uR}       {\mbox{\boldmath$R$}}

\newcommand{\uu}       {\mbox{\boldmath$u$}}
\newcommand{\uV}       {\mbox{\boldmath$V$}}

\newcommand{\uW}       {\mbox{\boldmath$W$}}

\newcommand{\uX}       {\mbox{\boldmath$X$}}

\newcommand{\uY}       {\mbox{\boldmath$Y$}}

\newcommand{\uZ}       {\mbox{\boldmath$Z$}}

\newcommand{\ualpha}            {\mbox{\boldmath$\alpha$}}
\newcommand{\ubeta}             {\mbox{\boldmath$\beta$}}
\newcommand{\ugamma}            {\mbox{\boldmath$\gamma$}}
\newcommand{\udelta}            {\mbox{\boldmath$\delta$}}
\newcommand{\uepsilon}          {\mbox{\boldmath$\epsilon$}}

\newcommand{\utheta}            {\mbox{\boldmath$\theta$}}
\newcommand{\uvartheta}         {\mbox{\boldmath$\vartheta$}}
\newcommand{\uiota}             {\mbox{\boldmath$\uiota$}}

\newcommand{\utau}              {\mbox{\boldmath$\tau$}}

\newcommand{\uphi}              {\mbox{\boldmath$\phi$}}

\newcommand{\uLambda}           {\mbox{\boldmath$\Lambda$}}

\newcommand{\uPi}               {\mbox{\boldmath$\Pi$}}
\newcommand{\uSigma}            {\mbox{\boldmath$\Sigma$}}

\newcommand{\uPhi}              {\mbox{\boldmath$\Phi$}}
\newcommand{\uPsi}              {\mbox{\boldmath$\Psi$}}

\newcommand{\sgn}{\mbox{$\text{sgn}$}}

\newcommand{\E}  {\mbox{E}}

\begin{document}
  \title{\bf Variable selection for varying multi-index coefficients models with applications to synergistic G$\times$E interactions}
\date{\vspace{-12ex}}
\maketitle
\begin{center}
  \author{Shunjie Guan$^{1}$\footnote{Presently at Pfizer Inc. The first two authors contributed equally to the work.}, Mingtao Zhao$^{2}$\footnote{Part of the work was done when the second author visited Michigan State University in 2019-2020.} and Yuehua Cui$^{1}$}\footnote{To whom the correspondence should be addressed: cuiy@msu.edu}\\
$^{1}${\it Department of Statistics and Probability, Michigan State University, East Lansing, MI 48824}\\
 $^{2}${\it School of Statistics and Applied Mathematics, Anhui University of Finance \& Economics, Bengbu, Anhui, 233000, China}\\
\end{center}

%
%
%




\begin{abstract}
Epidemiological evidence suggests that simultaneous exposures to multiple environmental risk factors (Es) can increase disease risk larger than the additive effect of individual exposure acting alone. The interaction between a gene and multiple Es on a disease risk is termed as synergistic gene-environment interactions (synG$\times$E). Varying multi-index coefficients models (VMICM) have been a promising tool to model synergistic G$\times$E effect and to understand how multiple Es jointly influence genetic risks on a disease outcome. In this work, we proposed a 3-step variable selection approach for VMICM to estimate different effects of gene variables: varying, non-zero constant and zero effects which respectively correspond to nonlinear synG$\times$E, no synG$\times$E and no genetic effect. For multiple environmental exposure variables, we also estimated and selected important environmental variables that contribute to the synergistic interaction effect. We theoretically evaluated the oracle property of the proposed variable selection approach. Extensive simulation studies were conducted to evaluate the finite sample performance of the method, considering both continuous and discrete gene variables. Application to a real dataset further demonstrated the utility of the method. Our method has broad applications in areas where the purpose is to identify synergistic interaction effect.
\end{abstract}


\textbf{Keyword}: Variable selection; Varying multi-index coefficients model; Synergistic G$\times$E interaction; Oracle property

\section{Introduction}
Genetic factors play fundamental roles in many complex diseases, and their effects are largely influenced by environmental factors. The same genetic factor can have different effects on disease risks under different environmental conditions, leading to the so called gene-environment (G$\times$E) interaction \cite{Falconer1952}. The identification of  G$\times$E interactions has been one of the central foci in genetic studies.

Recently, Ma et al. \cite{Ma2011} and Wu et al. \cite{Wu2013} proposed a nonparametric method to capture nonlinear G$\times$E interaction effects. Motivated by epidemiological evidence that simultaneously exposure to multiple environmental conditions would give rise to a higher risk than the simple addition of individual exposure acting alone, Liu et al. \cite{Liu2016} proposed a partial linear varying multi-index coefficients model to capture the interaction effect between genetic factors and multiple exposures, termed as synergistic G$\times$E (synG$\times$E). The method can test the interaction between a gene and a mixture of environmental variables and further assess if the interaction effect is linear or nonlinear. While the method was proposed under a low dimensional framework, when the number of genetic variables is large, a high-dimensional variable selection method is needed.

Consider the following varying multi-index coefficient model (VMICM),
\begin{equation}
\label{sivcm}
Y = \uf^T( \uX^T \ubeta )\uG + \epsilon,
\end{equation}
where $Y$ is a continuous response variable that measures certain phenotypic trait of interest;
$\uX \in \mathbb{R}^q$ is a $q$ dimensional environmental exposure variable and also called loading covariates;
$\uG\in \mathbb{R}^{p+1}$ is a $p+1$ dimensional genetic variable; $\uf(\cdot)=(f_0(\cdot),f_1(\cdot),\ldots,f_p(\cdot))^T$ is a $(p+1)\times 1$ vector of unknown functions with $f_k(\cdot)~(k=0,1,\ldots,p)$ being the $k$th unknown non-parametric function;
$\ubeta=(\beta_1,\beta_2,\ldots,\beta_q)^T$ is a vector of unknown loading parameters of dimension $q$.
The model error $\epsilon$ has mean $0$ and finite variance $\sigma^2$.
Furthermore, for the sake of identifiability, we assume $\|\ubeta\|=1$, $\beta_1 > 0$, where $\|\cdot\|$ denotes the Euclidean norm operator; and $f_k(\cdot)~(k=0,1,\ldots,p)$ cannot be the form as
  $\uf(\uu) = \ualpha^T\uu\ubeta^T\uu + \ugamma^T\uu + c_0$, where
 $\ualpha, \ugamma\in \mathbb{R}^{p+1}, c_0\in \mathbb{R}$ are constants, and $\ualpha$ and $\ubeta$ are not parallel to each other.

One of the main advantages of model (\ref{sivcm}) is that it models the effects of $\uG$ on $Y$ as functions of $\uX$ without suffering the curse of dimensionality.
One can interpret $f_k(\uX^T\ubeta)$ as the effect of $\uG$ on $Y$, modified by multiple $X$ variables through the index $\uX^T\ubeta$. In addition, model (\ref{sivcm}) is very flexible to cover a wide range of models. For instance, if $q=1$ and $\ubeta = 1$, then it becomes a varying-coefficient model; and if $p = 0$ and $\uG = 1$, then it becomes a standard single-index model.

Variable selection has been a popular statistical strategy to solve large $p$ small $n$ problems in a regression setup. In the past, researchers often opted for forward/backward selection, as well as information based criteria such as AIC and BIC for variable selection. Recently, variable selection via penalized regression has been gaining more popularity since it features simultaneous selection and estimation of parameters. The idea is to add a penalty function to the loss function or log-likelihood function. Bridge regression \cite{Frank1993}, least absolute shrinkage and selection operator (LASSO) \cite{Tibshirani1996} and its extensions (adaptive-LASSO\cite{Zou2006}), smoothly clipped absolute deviation (SCAD)\cite{Fan2001} and minimax concave penalty (MCP)\cite{Zhang2010} are a few examples. To evaluate different penalized functions, Fan and Li \cite{Fan2001} proposed three important criteria: sparsity, unbiasedness and continuity. They showed that SCAD penalty possess the oracle property, meaning that penalized regression featuring SCAD works as well as if the correct sub-model was known in advance. Adaptive LASSO\cite{Zou2006}, SCAD\cite{Fan2001} and MCP\cite{Zhang2010} all possess the oracle property. However, for adaptive LASSO, determining weights for parameters might become problematic when the dimension of a model is higher than sample size. In the current work, we applied MCP penalty function for its oracle property and fast algorithm.

Considering the complicated structure of model (\ref{sivcm}), specifically, the nonlinear structure about the unknown non-parametric functions $f_k(\cdot)~(k=0,1,\ldots,p)$ and the unknown parameter $\ubeta$, we proposed a three stage iterative variable selection strategy. Specifically, our goal is to: (1) classify the non-parametric functions $f_k(\cdot)~(k = 1,\ldots,p)$ into three categories: varying, non-zero constant and zero; (2) select zero and non-zero component of loading parameters $\ubeta$; and (3) estimate $f_k(\cdot)~(k = 0,1,\ldots,p)$ and $\ubeta$. Our approach was motivated by the practical need to separate three different mechanisms in G$\times$E interaction. The zero function of $f_k(\cdot)~(k=1,\ldots,p)$ indicates no genetic effect at all; the constant function of $f_k(\cdot)~(k=1,\ldots,p)$ indicates the effect of $G_k$ on $Y$ does not change over $\uX^T\ubeta$, hence no G$\times$E effect; while the varying function of $f_k(\cdot)~(k=1,\ldots,p)$ indicates the existence of G$\times$E effect. In addition to the selection of the coefficient functions, we can also select important loading parameters inside each index coefficient function, to further quantify the relative importance of individual exposure variables. If more than one $\uX$ variable is selected, we can conclude there is synG$\times$E effect. As shown in Liu et al. \cite{Liu2016}, the model has the advantage to capture the joint interaction of a gene with multiple exposures as a whole. Novel insights about the underlying genetic mechanism can be revealed by the proposed model.

Feng and Xue \cite{Feng2013} proposed a variable selection approach based on model (\ref{sivcm}) by applying a group SCAD penalty on B-spline coefficients and loading parameters $\ubeta$.
They focused on either zero or non-zero coefficient functions $f_k(\cdot)~(k = 0,1,\ldots,p)$.
We are particularly interested in the constant coefficient since it corresponds to no G$\times$E effect and has important practical implications.
Tang et al. \cite{Tang2012} and Wu et al. \cite{Wu2018} proposed a two step variable selection approach based on an additive varying-coefficient model.
They classified the non-parametric function into three categories: varying, constant or zero. Their model is a special case of our VMICM model when the dimension of the $\uX$ variable is one.
No variable selection approach on VMICM has been proposed to classify unknown non-parametric functions $f_k(\cdot)~(k = 1,\ldots,p)$ into three categories (varying, constant or zero), while selecting non-zero loading parameter $\ubeta$ simultaneously.
Following the previous work, we used B-spline basis functions to approximate unknown non-parametric functions $f_k(\cdot)~(k = 0,1,\ldots,p)$, then using penalized regression to classify $f_k(\cdot)~(k = 0,1,\ldots,p)$ into varying, constant or zero.
Further, we selected non-zero $\ubeta$ via first order approximation and penalized regression.
We showed that under some mild regularity conditions, our estimators possess the oracle property, indicating that our penalized estimators work as well as if the correct sub-model is known in advance.

The rest of the paper was organized as follows. Section 2 introduced our proposed variable selection approach, including the iterative estimation approach and how to select various tuning parameters for B-spline approximation and penalized regressions. Method on how to select initial values for $\ubeta$ was discussed. In Section 3, we evaluated the theoretical properties of our approach. In Section 4, we performed simulations to evaluate the performance of our method in finite samples, followed by a real data application in Section 5 and a discussion in Section 6.

\section{The variable selection method}

\subsection{Model setup}
Consider model (\ref{sivcm}) with data $\{(Y_i,\uX_i,G_{ik}), i=1,2,\ldots n, k=0, 1,2,\ldots, p\}$ in the following form,
\begin{equation}\label{VMICM}
  Y_i =  \uf^T( \uX_i^T \ubeta )\uG_i + \epsilon_i,~~i=1,2,\ldots,n,
\end{equation}
where $Y_i$ is a continuous response variable; $\uX_i = (X_{i1}, X_{i2}, \ldots, X_{iq})^T$ is $q$-dimensional continuous loading covariates;
$\uX_i^T\ubeta$ is the so-called index;
$\uG_i=(G_{ik})_{(p+1)\times n}=(\uG_1, \uG_2, \ldots, \uG_n)$,
$\uG_i = (1, G_{i1}, \ldots, G_{ip})^T$; $\uG_{\cdot k}=(\uG_{1k}, \ldots, \uG_{nk})^T$ is a continuous or discrete vector of length $n$ for $k = 0, 1, 2, \ldots p$.
In model (\ref{sivcm}), $f_k( \cdot )$ is the effect of $G_{\cdot k}$ on $Y$ for $k\neq 0$ and $f_0( \cdot )$ is the intercept function which models the marginal effect of $\uX$ on $Y$;
$\epsilon_i~(i=1, 2, \ldots, n)$ are unknown random errors with mean 0 and finite variance $\sigma^2$.
We further assume that $\epsilon_i$ and $\epsilon_j$ are independent of each other for $i\neq j$ ($\forall\ 1\leq i,j\leq n$), $\{\epsilon_i,i=1,2,\ldots,n\}$ are independent of $\{(\uX_i, G_{ik}),i=1,2,\ldots,n, k=1,2,\ldots, p\}$.

\subsection{Estimation method}

We approximate the unknown functions $\{f_k(u):u\in\mathcal{U}\}~(k=0,1,2,\ldots,p)$ using B-spline basis functions.
Here, we assume that $\mathcal{U}$ is a nondegenerate compact interval.
Denote $\mathscr{F}$ to be a collection of functions $f(u)$ satisfying (A2) in Appendix.
Let $K$ be the number of interior knots and $h$ be the order of the B-spline basis function.
By Schumaker (1981, chapter 4)\cite{Schumaker1981}, we can normalize the B-spline basis function $\widetilde{\uB}(u) = (\tilde{B}_1(u) , \tilde{B}_2(u),\ldots, \tilde{B}_L(u))^T$ for $\mathscr{F}$, and there exists a linear transformation matrix $\uPi$ \cite{Tang2012}, such that
\begin{equation}\label{Bspl}
  \uPi\widetilde{\uB}(u) = (\bm{1}, B_2(u), B_3(u),\ldots,B_L(u))^T = (\bm{1},\bar{\uB}^T(u))^T \buildrel \Delta \over =\uB(u)
\end{equation}
where $\bar{\uB}(u)= (B_2(u), B_3(u),\ldots, B_L(u))^T$, $L=K+h$ and each component of $\bar{\uB}(u)$ and $\widetilde{\uB}(u)$ is a function of $u$.
Clearly, $\uB(u)$ is also a basis function for $\mathscr{F}$.
In our work, we assume that $f_k(u)\in \mathscr{F}$ for $k=0,1,2,\ldots,p$.
Therefore, we can approximate each $f_k(u)$ by
\begin{equation}\label{Bsplpprox}
  f_k(u) \approx \uB^T(u)\ugamma_k = \gamma_{k1} + \bar{\uB}^T(u)\ugamma_{k*},~~k=0,1,2,\ldots,p,
\end{equation}
where $\ugamma_k = (\gamma_{k1},\ugamma_{k*}^T)^T $ and $\gamma_{k1}$ corresponds to the constant part of the coefficient function and $\ugamma_{k*} = (\gamma_{k2},\gamma_{k3},\ldots,\gamma_{kL})^T$ corresponds to the varying part.
To fix notation, we take $\ugamma=(\ugamma_0^T,\ugamma_1^T,\ldots, \ugamma_p^T)^T$, $\uW_i(\ubeta)= \uI_{p+1}\otimes \uB(\uX_i^T\ubeta)\cdot \uG_{i}$, where ${\uI_{p+1}}$ is the  $ (p+1) \times (p+1) $ identity matrix and ``$ \otimes$" is the Kronecker product operator.
With the B-spline approximation same as (\ref{Bsplpprox}), model (\ref{VMICM}) can be rewritten as
\begin{equation}\label{VMICMapprox1}
  Y_i \approx  \uW_i^T(\ubeta)\ugamma + \epsilon_i, ~~~~i=1,2,\ldots,n.
\end{equation}
In matrix notation, we have
\begin{equation}\label{VMICMapprox2}
  \uY \approx \uW(\ubeta)\ugamma + \uepsilon
\end{equation}
where $\uepsilon=(\epsilon_1, \epsilon_2,\ldots, \epsilon_n)^T$ and $\uW(\ubeta)=(\uW_1(\ubeta), \uW_2(\ubeta), \ldots,\uW_n(\ubeta))^T \in \mathbb{R}^n\times \mathbb{R}^{L(p+1)}$.
Thus, the original estimation problem can be transformed to estimate $\ugamma$ and $\ubeta$.

{\bf Remark 1}: By some simple matrix calculation, we can see that
\begin{equation}
  \uW_i^T(\ubeta)\ugamma = G_i^T \ugamma_{*1} + \bar{\uW}(\ubeta)_i^T\ugamma_*, ~~~~i=1,2,\ldots,n,
\end{equation}
where $\bar{\uW}^T_i(\ubeta)= \uI_{p+1}\otimes \bar{\uB}(\uX_i^T\ubeta)\cdot \uG_{i}$, $\ugamma_{*1} = (\gamma_{01}, \gamma_{11}, \ldots, \gamma_{p1})^T$ and $\ugamma_*=(\ugamma_{0*}^T, \ugamma_{1*}^T, \ldots, \ugamma_{p*}^T)^T$.

{\bf Remark 2}: The transformation matrix $\uPi$ can separate the main genetic and G$\times$E effect from the total effect, which further enables us to assess if there exist genetic main and interaction effects, that is: (1) if $\|\ugamma_{k*}\| =(\sum_{l=2}^L \gamma_{kl}^2)^{1/2}\neq 0$, then there exists interaction between $G_{\cdot k}$ and multiple $\uX$; (2) if $\|\ugamma_{k*}\| = 0$ and $|\gamma_{k1}|\neq 0$, then $G_{\cdot k}$ has a constant effect on $Y$, i.e., no G$\times$E interaction effect; and (3) if further $\|\ugamma_{k*}\| = 0$ and $|\gamma_{k1}| = 0$ then $G_{\cdot k}$ has no effect on $Y$ at all.

To select and estimate the parameters $\ugamma$ and $\ubeta$, we apply the penalized regression idea and minimize the following penalized least squares objective function
\begin{equation}\label{Q}
  \begin{split}
Q(\ubeta,\ugamma) =& \sum_{i=1}^n \left(Y_i -\uW_i^T(\ubeta)\ugamma\right)^2 + n\sum_{k=1}^p p_{\lambda_{1k}}(\|\ugamma_{k*}\|)  \\
& + n\sum_{k=1}^p p_{\lambda_{2k}}(|\gamma_{k1}|)I(\|\ugamma_{k*}\| = 0) + n\sum_{d=2}^q p_{\lambda_{3d}}(|\beta_{d}|),
\end{split}
\end{equation}
where $p_{\lambda_{1k}}(\cdot), p_{\lambda_{2k}}(\cdot), p_{\lambda_{3d}}(\cdot)$ are penalty functions of the corresponding parameters, and $I(\cdot)$ is an indicator function.
In our work, the penalty functions $p_{\lambda_{1k}}(\cdot), p_{\lambda_{2k}}(\cdot), p_{\lambda_{3d}}(\cdot)$ are MCP \cite{Zhang2010} penalty functions such that $p(x,\lambda) = \lambda \int_0^x(1-\frac{s}{\tau\lambda})_+ds$ with regularization parameters $\tau>0$ and $\lambda>0$.

{\bf Remark 3} : (1) From the construction of the penalty function, we penalize $\gamma_{k1}$ only if $\|\ugamma_{k*}\| = 0$. If $\|\ugamma_{k*}\| \neq 0$, it implies that the function is varying and no need to penalize the constant part;
(2) No penalty is applied to the intercept function $f_0(\cdot)$. There is no practical motivation to penalize the marginal intercept function; and
(3) No penalty is applied to the first loading parameter $\beta_{1}$ in $\ubeta$ due to the constraint.

We now handle the constraints $\|\ubeta\|=1$ and $\beta_1 > 0$ on the $q$-dimensional single-index parameter $\ubeta$ with reparametrization.
Denote $\uphi=(\phi_2,\phi_3,\ldots, \phi_q)^T=(\beta_2,\beta_3,\ldots,\beta_q )^T$, and we can get $$\ubeta = \left(\sqrt{1-\|\uphi\|^2}, \phi^T\right)^T, ~~~ \|\uphi\|<1.$$
Therefore, $\ubeta = \ubeta(\phi)$, and $\ubeta$ is infinitely differentiable with respect to $\uphi$.
The Jacobian matrix of $\ubeta$ with respect to $\uphi$ is

\begin{equation}\label{JacobM}
\uJ_{\uphi}  = \left( {\begin{array}{*{20}{c}}
{ - {{(1 - {{\| \uphi  \|}^2})}^{ - 1/2}}{\uphi ^T}}\\
\uI_{q-1}
\end{array}} \right).
\end{equation}
Note that $\uphi$ is one dimension lower than $\ubeta$, and $Q(\ubeta, \ugamma)$ can be rewritten as
\begin{equation}\label{Qphi}
  \begin{split}
Q(\uphi,\ugamma) =& \sum_{i=1}^n \left(Y_i -\uW_i^T(\uphi)\ugamma \right)^2 + n\sum_{k=1}^p p_{\lambda_{1k}}(\|\ugamma_{k*}\|)  \\
& + n\sum_{k=1}^p p_{\lambda_{2k}}(|\gamma_{k1}|)I(\|\ugamma_{k*}\| = 0) + n\sum_{d=2}^q p_{\lambda_{3d}}(|\phi_d|),
\end{split}
\end{equation}
where $\uW_i(\uphi) = \uW_i(\ubeta)$. Then we can get the penalized least squares estimators $\hat \uphi$, $\hat \ugamma$ and $\hat \ubeta$ as
\begin{equation}\label{estimators}
 (\hat{\uphi}, \hat\ugamma) = \arg\min_{\uphi, \ugamma}Q(\uphi,\ugamma),
\end{equation}
\begin{equation}\label{betahat}
  \hat\ubeta = \left(\sqrt{1-\|\hat\uphi\|^2}, \hat\uphi^T\right)^T,~~\|\hat\uphi\|\leq 1.
\end{equation}
where $\hat\ugamma=(\hat\ugamma_0^T, \hat\ugamma_1^T, \ldots, \hat\ugamma_p^T)^T$.
Therefore, the estimator of $f_k(u)$ can be obtained by
\begin{equation}\label{mkhat}
  \hat{f}_k(u) = \uB^T(u)\hat{\gamma}_k, ~~~~ k=0,1,2,\ldots, p.
\end{equation}



\subsection{Iterative algorithm}
We can see that $\hat{\uphi}$ and $ \hat\ugamma$ denoted by (\ref{estimators}) do not have closed form.
Thus, we propose a iterative approach to get the numerical solution of $\hat{\uphi}$ and $ \hat\ugamma$.
Our modeling purpose is to separate $f_k(\cdot)~(k=1,2,\ldots, p)$ into three different categories: varying, non-zero constant or zero, denoted by $\mathcal{V}$, $\mathcal{C}$ and $\mathcal{Z}$ respectively.
For $ \forall k \in \{1, 2, \ldots, p\}$, notations ``$k \in \mathcal{V}$", ``$k \in \mathcal{C}$" and ``$k \in \mathcal{Z}$" mean that the function $f_k(\cdot)$ is varying, non-zero constant and zero respectively.
Obviously, $\mathcal{V}, \mathcal{C}$ and $\mathcal{Z}$ are mutually disjoint, and $\mathcal{V}\cup\mathcal{C}\cup\mathcal{Z}=\{1,2,\ldots,p\}$.
Furthermore, $k \notin \mathcal{V}$ means that $f_k(\cdot)$ is non-zero constant or zero, that is, $\{k \notin \mathcal{V}\}=\{k \in \mathcal{C}\}\cup \{k \in \mathcal{Z}\}$.
Following Feng and Xue \cite{Feng2013} and Tang et al. \cite{Tang2012}, we propose a stepwise iterative approach to obtain our penalized estimator.

\textbf{Step 0}:
Set initial values $\hat{\ubeta}^{(0)}$ and $\hat\ugamma^{(0)}$ to start the iteration.
Setting $f_k(\cdot)~(k=0,1,2,\ldots,p)$ as identity functions, we can get a simple linear additive model as
\begin{equation}\label{slam}
  Y_i = \uX_i^T\ubeta + \uX_i^T\ubeta \cdot G_{i1} + \uX_i^T\ubeta \cdot G_{i2}+ \ldots + \uX_i^T\ubeta \cdot G_{ip} + \epsilon_i,~~i=1,2,\ldots,n.
\end{equation}
Therefore, we can set an initial estimator $\tilde\ubeta= (\tilde\beta_1, \tilde{\uphi}^T)^T$ as
\begin{equation}\label{betatilde}
  \tilde{\ubeta} = (\tilde{\uX}^T\tilde{\uX})^{-1}\tilde{\uX}^TY,
\end{equation}
where $\tilde\uphi=(\tilde\beta_2, \tilde\beta_3, \ldots, \tilde\beta_q)^T$, $\tilde{\uX} = (\uX_1, \tilde{G}_2\uX_2,\ldots,\tilde{G}_n\uX_n)^T$, $\tilde{G}_i = \sum_{k=1}^{p}G_{ik}$.
Considering the constraints for $\ubeta$ such that $\|\ubeta\|=1$ and $\beta_1 > 0$, the initial estimator $\hat \ubeta^{(0)}$ can be chosen from (\ref{slam}) and (\ref{betatilde}) as
\begin{equation}\label{betahat0}
  \hat\ubeta^{(0)}=\frac{\tilde\ubeta}{\|\tilde\ubeta\|}\cdot \sgn(\tilde{\beta}_1)
\end{equation}
Then the initial estimator of $\hat{\ugamma}^{(0)}$ can be obtained by
\begin{equation}\label{gammahat0}
  \hat{\ugamma}^{(0)}=\left( \sum_{i=1}^{n}\uW_i(\hat \ubeta^{(0)})\uW_i^T(\hat \ubeta^{(0)})\right)^{-1}
  \sum_{i=1}^{n}\uW_i^T(\hat \ubeta^{(0)}) Y_i.
\end{equation}



\textbf{Step 1}:
In this step, we classify $f_k(\cdot)~(k=1,2,\ldots,p)$ into varying ($k\in\mathcal{V}$) and non-varying ($k\in\mathcal{C}\cup\mathcal{Z}$).
For a given initial value of $\ubeta$, denoted by $\hat{\ubeta}^{(0)}$ from (\ref{betahat0}), we can obtain our 1st step estimation $\hat{\ugamma}^{(1)}=((\hat{\ugamma}^{(1)}_0)^T, (\hat{\ugamma}^{(1)}_1)^T, \cdots, (\hat{\ugamma}^{(1)}_p)^T)^T$ by following a group penalized regression
\begin{equation}\label{gammahat(1)}
  \hat{\ugamma}^{(1)}  = \min_{\ugamma} Q_1(\ugamma|\Lambda_1,\hat{\ubeta}^{(0)}),
\end{equation}
where the $k$th coefficient $\hat{\ugamma}^{(1)}_k = (\hat{\gamma}_{k1}^{(1)},(\hat{\ugamma}_{k*}^{(1)})^T)^T~(k=0,1,2,\ldots, p)$,
$\Lambda_1=\{\lambda_{11},\lambda_{12},\ldots,\lambda_{1p}\}$ and
\begin{equation}\label{Q1}
  Q_1(\ugamma|\Lambda_1,\hat{\ubeta}^{(0)}) = \sum_{i=1}^n \left(Y_i -\uW_i^T(\hat{\ubeta}^{(0)})\ugamma\right)^2+ n\sum_{k=1}^p p_{\lambda_{1k}}(\|\ugamma_{k*}\|).
\end{equation}

Note that $\|\ugamma_{k*}\| > 0$ and $\|\ugamma_{k*}\| = 0$ respectively imply that $f_k(\cdot)$ is varying ($k\in\mathcal{V}$) and non-varying ($k \in \mathcal{C}\cup\mathcal{Z}$).
Therefore, instead of penalizing each coordinate of $\ugamma_{k*} = (\gamma_{k2},\ldots,\gamma_{kL})^T~(k=1,2,\ldots,p)$ separately, we penalized  $\|\ugamma_{k*}\|~(k=1,2,\ldots,p)$ for the reason that we want to assess the presence of the joint varying effect of $\uX$ and $G_{\cdot k}$ on $Y$.
In particular, from (\ref{Q1}), no penalty is applied to $\ugamma_{0*}$, which means that the intercept function $f_0(\cdot)$ is treated as being varying in our work.
Step 1 separates $f_k(\cdot)~(k = 1,\ldots,p)$ into two categories, i.e., varying and non-varying.
However, $\hat{\ugamma}^{(1)}$ does not have a closed form. We can only get numerical solutions through an iterative algorithm.
The detailed iterative algorithm for this step can be found in A.1 of the Appendix, with the initial iterative value of $\ugamma$ denoted by $\hat{\ugamma}^{(0)}$ in (\ref{gammahat0}).

\textbf{Step 2}: After Step 1, we would like to further select variables with constant effects and separate the non-varying functions $f_k(\cdot)~(k\in \mathcal{C}\cup\mathcal{Z})$ into non-zero constants ($k\in\mathcal{C}$) and zeros ($k\in\mathcal{Z}$) in this step, i.e.,
estimate and select $\gamma_{k1}$ given $\hat{\ugamma}_{k*}^{(1)} = 0$ for $k\in\mathcal{C}\cup\mathcal{Z}$.
In order to do that, we penalize $\gamma_{k1}$ only when $\|\hat{\ugamma}_{k*}^{(1)}\| = 0$, i.e. $k\in\mathcal{C}\cup\mathcal{Z}$, and no penalty is applied to $\gamma_{01}$.

We obtain estimator $\hat{\ugamma}^{(2)}=((\hat{\ugamma}^{(2)}_0)^T, (\hat{\ugamma}^{(2)}_1)^T, \cdots, (\hat{\ugamma}^{(2)}_p)^T )^T$ via penalized regression
\begin{equation}\label{gammahat(2)}
\hat{\ugamma}^{(2)} = \min_{\ugamma}Q_2(\ugamma|\Lambda_2,\hat{\ubeta}^{(0)},\hat{\ugamma}^{(1)}),
\end{equation}
where $(\hat{\ugamma}_k^{(2)})_{k\in \mathcal{V}} = (\hat{\gamma}_{k1}^{(2)},(\hat{\ugamma}_{k*}^{(2)})^T)^T, (\hat{\ugamma}_k^{(2)})_{k\in \mathcal{C}} = (\hat{\gamma}_{k1}^{(2)},\textbf{0}^T)^T, (\hat{\ugamma}_k^{(2)})_{k\in \mathcal{Z}} = \textbf{0},k=1,2,\ldots,p$,
$\Lambda_2=\{\lambda_{21},\lambda_{22},\ldots,\lambda_{2p}\}$ and
\begin{equation}\label{Q2}
  \begin{split}
     Q_2(\ugamma|\Lambda_2,\hat\ubeta^{(0)},\hat{\ugamma}^{(1)})
     &= \sum_{i=1}^n \left(Y_i -\uW_i^T(\hat{\ubeta}^{(0)})\hat{\ugamma}^{(1)}\right)^2    + n\sum_{k=1}^p p_{\lambda_{2k}}(|\gamma_{k1}^{(1)}|)I(\|\hat{\ugamma}_{k*}^{(1)}\| = 0).
  \end{split}
\end{equation}
The detailed iterative algorithm for Step 2 can be found in A.1 of the Appendix.
After Step 1 and 2, we can obtain the estimator of the B-spline coefficients $\ugamma$ denoted as $\hat{\ugamma}^{(2)}$ and separate $f_k(\cdot)~(k=1,\ldots,p)$ into $\mathcal{V}$, $\mathcal{C}$ or $\mathcal{Z}$.
Then the next step is to estimate and select loading parameter $\ubeta$ given $\hat{\ugamma}^{(2)}$.

\textbf{Step 3}: We obtain $\hat{\ubeta}$ via the penalized regression
\begin{equation}\label{betahat}
  \hat{\ubeta} = \min_{\|\ubeta\|=1}Q_3(\ubeta|\Lambda_3, \hat{\ugamma}^{(2)}),
\end{equation}
where $\Lambda_3=\{\lambda_{32},\ldots,\lambda_{3q}\}$ and
\begin{equation}\label{Q3}
  Q_3(\ubeta|\Lambda_3, \hat{\ugamma}^{(2)}) = \sum_{i=1}^n \left(Y_i -\uW_i^T(\ubeta)\hat{\ugamma}^{(2)}\right)^2 + n\sum_{d=2}^{q} p_{\lambda_{3d}}(|\phi_d|).
\end{equation}
The detailed iterative algorithm for $\hat{\ubeta}$ can be found in A.1 of the Appendix.
We then replace $\hat{\ubeta}^{(0)}$ by $\hat{\ubeta}$ and iterate between Step 1 and Step 3 until convergence.


\subsection{Selection of tuning parameters}
We use the Bayesian Information Criterion (BIC) to select the tuning parameters $\tau$, $\lambda_{1k}$, $\lambda_{2k}$ and $\lambda_{3d}$ in the penalty functions\cite{Schwarz1978}. Since there are too many tuning parameters in our penalty functions, and the minimization problem for the BIC method over a high-dimensional space is computationally intensive and difficult to track, similar to Feng and Xue \cite{Feng2013}, we take the adaptive tuning parameters $\lambda_{1k}$, $\lambda_{2k}$ and $\lambda_{3d}$ as
\begin{equation*}
  \lambda_{1k} = \frac{\lambda_1}{\|\hat\ugamma_k^{un}\|},~\lambda_{2k} = \frac{\lambda_2}{\|\hat\gamma_{k1}^{(1)}\|},~
    \lambda_{3d} = \frac{\lambda_3}{|\hat\beta_d^{un}|}
\end{equation*}
where $\hat\ugamma_k^{un}~(k=1,2,\ldots,p)$ and $\hat\beta_d^{un}~(d=2,\ldots,q)$ are the unpenalized estimators of $\ugamma_k~(k=1,2,\ldots,p)$ and $\beta_d^{(0)}~(d=2,\ldots,q)$.
$\hat\ugamma_k^{(1)}=(\gamma_{k1}^{(1)}, (\hat\ugamma_{k*}^{(1)})^T)^T$ is denoted by (\ref{gammahat(1)}) and satisfies $\|\hat\ugamma_{k*}^{(1)}\|=0$.
Therefore, we transform the selection of tuning parameters $\lambda_{1k}$, $\lambda_{2k}$ and $\lambda_{3d}$ into a one-dimensional grid searching problem.
We just need to chose optimal $\lambda_1$, $\lambda_2$ and $\lambda_3$ in the three step algorithm.

In Step 1, we take optimal $\lambda_1$ as the minimizer of
\begin{equation}\label{BIC1}
   \text{BIC}_1(\lambda_1)= \log\left( \sum_{i=1}^n \left(Y_i - \uW_i^T(\hat{\ubeta}^{(0)})\hat{\ugamma}^{(\lambda_1)}\right )^2\right) + \frac{\log(n)}{n}\cdot df_{\lambda_1},
\end{equation}
where $\hat{\ugamma}^{(\lambda_1)} =\arg\mathop{\min}\limits_{\ugamma} Q_1(\ugamma|\Lambda_1,\hat{\ubeta}^{(0)}) $ is defined by (\ref{gammahat(1)}) for a given $\lambda_1$, $\hat{\ubeta}^{(0)}$ is denoted as (\ref{betahat0}), $df_{\lambda_1}$ is defined as the total number of non-zero coefficients of $\{\|\ugamma_k^{(\lambda_1)}\|, k=1,2,\ldots,p\}$ for a given $\lambda_1$.

In Step 2, the optimal $\lambda_2$ is the minimizer of
\begin{equation}\label{BIC2}
  \text{BIC}_2(\lambda_2) = \log\left( \sum_{i=1}^n \left(Y_i - \uW_i^T(\hat{\ubeta}^{(0)})\hat{\ugamma}^{(\lambda_2)}\right )^2 \right) + \frac{\log (n)}{n}\cdot df_{\lambda_2},
\end{equation}
where $\hat{\ugamma}^{(\lambda_2)}= \arg \mathop{\min}\limits_{\ugamma} Q_2(\ugamma|\Lambda_2,\hat{\ubeta}^{(0)}, \hat{\ugamma}^{(1)}) $ is defined by (\ref{gammahat(2)}) for a given $\lambda_2$,
$df_{\lambda_2}$ is defined as the total number of non zero coefficients of $\{\|\ugamma_k^{(\lambda_2)}\|, k=1,2,\ldots,p\}$ for a given $\lambda_2$.

In Step 3, we take optimal $\lambda_3$ as the minimizer of
\begin{equation}\label{BIC3}
  \text{BIC}(\lambda_3) = \log\left( \sum_{i=1}^n\left(Y_i - \uW_i^T(\hat{\ubeta}^{(\lambda_3)})\hat{\ugamma}^{(\lambda_2)}\right)^2 \right) + \frac{\log n}{n}\cdot df_{\lambda_3},
\end{equation}
where $\hat{\ugamma}^{(\lambda_2)}= \arg \mathop{\min}\limits_{\ugamma} Q_2(\ugamma|\Lambda_2,\hat{\ubeta}^{(0)}, \hat{\ugamma}^{(1)}) $, and $\hat{\ubeta}^{(\lambda_3)}=\arg\mathop{\min}\limits_{\ubeta}Q_3(\ubeta|\Lambda_3, \hat{\ugamma}^{(2)})$ is defined by (\ref{betahat}) for a given $\lambda_3$, and $df_{\lambda_3}$ is defined as the total number of non-zero $\beta_d~(d=1,2,\ldots,q)$ for a given $\lambda_3$.
We search the optimal value of $\lambda_1, \lambda_2, \lambda_3$ over a grid of 100 exponentially decreasing values with the minimum being 1E-3, and the maximum of $\lambda_1, \lambda_2, \lambda_3 $ is set to be the minimum value such that all of the penalized estimators are zeros.

\subsection{Selection of the order $h$ and the number of interior knots $K$}
Since $h$ is the order of the B-spline basis function, higher degree corresponds to more complicated interactions and is less interpretable in practice.
Tang et al.\cite{Tang2012} suggested using lower degree splines such as linear, quadratic or cubic splines.
For instance, $h =2, 3, 4$ represent linear, quadratic and cubic splines respectively.
Hence, we search optimal order $h_{opt}$ over the set $\mathcal{H}=\{2,3,4\}$.
Futhermore, $K = O_p(n^{\frac{1}{2r+1}})$ is a necessary assumption for oracle properties of the proposed variable selection approach, where $n$ is the sample size and $r$ is defined in condition (A2) in Appendix.
According to He et al. \cite{He2005}, in our work, the range of the interior knots is taken to be $\mathcal{K}=\left[\max(\lfloor 0.5\cdot n^{\frac{1}{2r+1}}\rfloor,1),\lfloor 1.5 \cdot n^{\frac{1}{2r+1}}\rfloor\right]$, where $\lfloor x\rfloor$ denotes the integer part of $x$.

In theory, we can select the optimal order $h_{opt}$ and the number of interior knots $K_{opt}$ for each nonparametric function $f_k(\cdot)$.
However, this is practically infeasible due to the large searching space and the computational cost.
We assume that all the nonparametric functions share common $h$ and $K$.
Thus, $(K_{opt},h_{opt})$ can be achieved via a two-dimensional grid search for $(K_{opt},h_{opt}) \in \mathcal{K}\times\mathcal{H}$ focusing only on the intercept function by the following criterion
\begin{equation}\label{optimal(Kh)}
  (K_{opt},h_{opt})=\arg\min_{K,h}\left\{\log\left( \sum_{i = 1}^n \left(Y_i- \uW_i^T\hat\ugamma \right)^2\right) + \frac{\log(n)}{n}(K + h)\right\},
\end{equation}
where $\hat{\ugamma}=(\hat{\ugamma}_0^T, \uzero^T, \ldots, \uzero^T)^T$.

\section{Theoretical Properties}
We first fix some notations.
Let $\uf_0(\cdot)=(f_{00}(\cdot), f _{10}(\cdot), \ldots, f_{p0}(\cdot))^T$ and $\ubeta_0 = (\beta_{10}, \beta_{20}, \ldots, \beta_{q0})^T$ be the true value of $\uf(\cdot)$ and $\ubeta$  respectively,
and denote $\ugamma_0 = (\ugamma^{T}_{00}, \ugamma^{T}_{10}, \ldots, \ugamma^{T}_{p0})^T$ be the true value of the B-spline coefficient $\ugamma$, where $\ugamma_{k0} = (\gamma^0_{k1}, \ugamma^{0T}_{k*})^T$,
$\ugamma^0_{k*}=(\gamma^0_{k2}, \gamma^0_{k3}, \ldots, \gamma^0_{kL})^T$.
Without loss of generality, we assume $\beta_{d0} \neq 0$ for $d = 1,\ldots s$, $\beta_{d0} = 0$ for $d = s+1,\ldots q$; $f_{k0}(\cdot)$ is varying for $k = 0,1,\ldots,v$, $f_{k0}(\cdot)$ is non-zero constant for $k = v+1,\ldots,c$ and $f_{k0}(\cdot)$ is zero for $k = c+1,\ldots,p$.
Clearly, we can see that $\mathcal{V}=\{0,1,2,\ldots,v\}$ and $\mathcal{C}=\{v+1,v+2,\dots,c\}$, $\mathcal{Z}=\{c+1,c+2,\ldots, p\}$.
The following theorem gives the consistency of the penalized least square estimators.

\begin{theorem}\label{Th1}
  Suppose the regulatory conditions (A1) - (A8) in Appendix hold and the number of interior knots $K = O_p(n^{1/(2r+1)})$. Then\\
\indent(i)  $\| \hat{\ubeta} - \ubeta_0 \| = O_p( n^{-r/(2r+1)} + a_n)$;\\
\indent(ii) $\| \hat{f}_k(\cdot) - f_{k0}(\cdot) \| = O_p( n^{-r/(2r+1)} + a_n )$, $k = 0, 1,\ldots,p$;\\
where
$a_n = \mathop{\max}\limits_{k,l}\{ p'_{\lambda_{1k}}(\|\ugamma_{k*}^0\|), p'_{\lambda_{2k}}(|\gamma_{k1}^0|), p'_{\lambda_{3l}}(|\beta_{d0}|) , \ugamma_{k*}^0\neq 0, \gamma_{k1}^0 \neq 0, \beta_d^0\neq 0, k = 1,2,\ldots, p, d = 1,2,\ldots, q  \}$.
\end{theorem}

Furthermore, under some regularity conditions, we can demonstrate that the above consistent estimators possess the following sparsity properties.

\begin{theorem}\label{Th2}
  Suppose the regularity conditions (A1) - (A8) in Appendix hold and the number of interior knots $K = O_p(n^{1/(2r+1)})$. Let
$\lambda_{\max} = \max \{\lambda_{1k},\lambda_{2k},\lambda_{3d}, k=1,2,\ldots,p;d=2,\ldots,q\}$ and $\lambda_{\min} = \min \{\lambda_{1k},\lambda_{2k},\lambda_{3d},k=1,2,\ldots,p;d=2,\ldots,q\}$.
Suppose $\lambda_{max}\to 0$ and $n^{r/(2r+1)}\lambda_{min}\to \infty$ as $n\to \infty$. Then with probability approaching to $1$, $\hat{\ubeta}$ and $\hat{f}_k(\cdot)$  satisfy\\
\indent(i)   $\hat{\beta}_d = 0$ for $d = s+1,\ldots,q$;\\
\indent(ii)  $\hat{f}_k(\cdot) = c_k$ for $k = v+1,\ldots, c$, where $c_k$ is some non-zero constant;\\
\indent(iii) $\hat{f}_k(\cdot) = 0$ for $k = c+1,\ldots, p;$
\end{theorem}

Next, we show that the asymptotic normality of the non-zero coefficients $\beta$ and the spline coefficients $\ugamma$.
Obviously, if $\mathcal{C} \neq \varnothing$, model (\ref{VMICM}) degenerates into a partial linear single-index varying-coefficient model.
However, the true model is unknown in advance.
Without loss of generality, we treat all of functions $f_k(\cdot)~(k=1,2,\ldots, p)$ as being varying in advance,
then identify whether each $f_k(\cdot)$ is varying, non-zero constant or zero.
Denote
$$\ubeta^* = (\beta_1, \beta_2, \ldots, \beta_s)^T,~~ \uf^*(\cdot)=(\uf^{*T}_{(\mathcal{V})}(\cdot),\uf^{*T}_{(\mathcal{C})}(\cdot))^{T},$$
$$\uf^*_{(\mathcal{V})}(\cdot) = (f_0(\cdot),f_1(\cdot),\ldots, f_v(\cdot))^T,
~~ \uf^*_{(\mathcal{C})}(\cdot) = (f_{v+1}(\cdot), f_{v+2}(\cdot), \ldots, f_c(\cdot))^T,$$
and the corresponding covariates are denoted by $\uX^*, \uG^*_i =(\uG^{*T}_{(\mathcal{V})i}, \uG^{*T}_{(\mathcal{C})i})^T~(i=1,2,\ldots,n)$.
Let $\ubeta^*_0  = (\beta_{10}, \beta_{20}, \ldots, \beta_{s0})^T$ and $\uf^*_0(\cdot) =(\uf^{*T}_{(\mathcal{V})0}(\cdot), \uf^{*T}_{(\mathcal{C})0}(\cdot))^T$ to be the true values of $\ubeta^*$ and $\uf^*(\cdot)$, where $\uf^*_{(\mathcal{V})0}(\cdot)=(f_{00}(\cdot),f_{10}(\cdot), \ldots,f_{v0}(\cdot) )^T$,
$\uf^*_{(\mathcal{C})0}(\cdot)=(f_{(v+1)0}(\cdot),f_{(v+2)0}(\cdot), \ldots,f_{c0}(\cdot) )^T$.
Obviously, $f_{k0}(u)~(k=v+1,v+2,\ldots,c)$ are non-zero constants for $\forall u \in \mathcal{U}$.
Similarly, we have $\uphi^*$, $\uW_i^* = (\uW^T_{(\mathcal{V})i}, \uW^T_{(\mathcal{C})i})^T$ and $\ugamma^* = (\ugamma^{*T}_{(\mathcal{V})}, \ugamma^{*T}_{(\mathcal{C})})^T$, where
$$\uW_{(\mathcal{V})i} = \uI_{v+1}\otimes \uB(\uX_i^{*T}\ubeta^*)\cdot \uG^*_{(\mathcal{V})i},~~\uW_{(\mathcal{C})i} = \uI_{c-v}\otimes \uB(\uX_i^{*T}\ubeta^*)\cdot \uG^*_{(\mathcal{C})i},$$
$$\ugamma^*_{(\mathcal{V})} = (\ugamma_{0}^{T}, \ugamma_{1}^{T}, \ldots, \ugamma_{v}^{T})^T,~~~~
\ugamma^*_{(\mathcal{C})} = (\ugamma_{v+1}^{T}, \ugamma_{v+2}^{T}, \ldots, \ugamma_{c}^{T})^T.$$
Denote $\ugamma^*_{(\mathcal{V})0} = (\ugamma^T_{00}, \ugamma^T_{10}, \ldots, \ugamma^T_{v0})^T$, $\ugamma^*_{(\mathcal{C})0} = (\ugamma^T_{(v+1)0}, \ugamma^T_{(v+2)0}, \ldots, \ugamma^T_{c0})^T$
be the estimators of the B-spline approximation to $\uf^*_{(\mathcal{V})0}(\cdot)$ and $\uf^*_{(\mathcal{C})0}(\cdot)$, respectively.
We can see that $\ugamma_{k0} = (\gamma_{k1}, 0, 0, \ldots,0)^T$ for $k = v+1, v+2, \ldots, c$ and $\gamma^0_{k1}~(k = v+1, v+2, \ldots, c)$ are non-zero constants.
Furthermore, we have $\uf^*_{(\mathcal{C})0} = (\gamma^0_{(v+1)1}, \gamma^0_{(v+2)1}, \ldots, \gamma^0_{c1})^T$.
Denote $\uvartheta^* = (\ugamma^{*T}_{(\mathcal{C})}, \uphi^{*T})^T$. The corresponding estimator and true value of $\uvartheta^*$ are denoted by $\uvartheta_0^* = (\ugamma^{*T}_{(\mathcal{C})0}, \uphi_0^{*T})^T$ and $\hat\uvartheta^* = (\hat\ugamma^{*T}_{(\mathcal{C})}, \hat\uphi^{*T})^T$, respectively.
In addition, let
\begin{equation}\label{Sigma1}
  \Sigma_1 = \E(\uG^{*}_{(\mathcal{C})}\uG^{*T}_{(\mathcal{C})})-\E\{C_1(\uX^{*T}_i\ubeta^*)D^{-1}(\uX^{*T}_i\ubeta^*)C_1^T(\uX^{*T}_i\ubeta^*)\}
\end{equation}
\begin{equation}\label{Sigma2}
  \Sigma_2 = \E(\uV^*\uV^{*T})-\E\{C_2(\uX^{*T}_i\ubeta^*)D^{-1}(\uX^{*T}_i\ubeta^*)C_2^T(\uX^{*T}_i\ubeta^*)\}
\end{equation}
where $$\uV^*= \dot{\uf}^T(\uX^{*T}_i\ubeta^*)\uG^*\uX^*, ~~~~ D(u)=\E\{\uG^*_{(\mathcal{V})}\uG^{*T}_{(\mathcal{V})}| \uX^{*T}_i\ubeta^* = u\}$$
$$C_1(u)=\E\{\uG^{*T}_{(\mathcal{C})}\uG^{*T}_{(\mathcal{V})}| \uX^{*T}_i\ubeta^* = u\}, ~~~~ C_2(u)=\E\{\uV^*\uG^{*T}_{(\mathcal{V})}| \uX^{*T}_i\ubeta^* = u\}$$
Then, we can get the asymptotic normality of $\hat{\uvartheta}^*$ in the following theorem.

\begin{theorem}
  Under the assumptions of Theorem 2, $\hat{\uvartheta}^*$ is $\sqrt{n}$-consistent and
  \begin{equation}\label{Th3}
    \sqrt{n}(\hat\uvartheta^* - \uvartheta^*) \xrightarrow{\mathscr{D}} N(0, \uSigma)
  \end{equation}
\end{theorem}
where notation ``$\xrightarrow{\mathscr{D}}$" represents ``convergence in distribution" and
$$\uSigma = \left( {\begin{array}{*{20}{c}}
\Sigma_1^{-1}&\uzero\\
\uzero& \uJ_{\uphi_0^*}\Sigma_2^{-1}\uJ^T_{\uphi_0^*}
\end{array}} \right).$$
All the proofs can be found in Appendix.

\section{Simulation}
We conducted extensive simulations to evaluate the performance of the proposed approach. The performance is measured in several ways: (1) classification accuracy of the $f(\cdot)$ function denoted as the oracle percentage; (2) IMSE of the estimated $f$-function; (3) selection accuracy of $\beta$; and (4) estimation accuracy of $\ubeta$ by MSE. Denote $R$ as the total number of simulation runs.

Oracle percentage of $f(\cdot)$ is defined as the percentage of correct classification out of a total of R simulations, for example, if $k\in \mathcal{V}$, and out of R simulations, $f_k(\cdot)$ is classified as varying for $g$ times, then the oracle percentage of $f_k(\cdot)$ is $\frac{g}{R} \times 100\%$.
IMSE of $f_k(\cdot)$ is defined as
\begin{equation}\label{IMSE}
  \text{IMSE}=\frac{1}{R} \sum_{\ell=1}^R\left( \frac{1}{n_{grid}} \sum_{j = 1}^{n_{grid}} \left(f_k(u_j)-\uB^T(u_j)\hat{\gamma}^{(\ell)}_k\right)^2 \right)
\end{equation}
where $n_{grid}$ is the number of points used to estimate the IMSE of the predicted function; $\hat{\gamma}_k^{(\ell)}$ are the estimators of the B-spline coefficients for the $\ell$th simulation; $\hat{\ubeta}^{(\ell)}$ is the estimator of the loading parameter $\ubeta$ for the $\ell$th simulation; $u_j$ is taken at the $j/n_{grid}\times 100\%$ quantile among the range of $\uX^T\hat{\ubeta}^{(\ell)}$. For our simulations, $n_{grid}$ was set to be $100$.

Oracle percentage of $\ubeta$ is defined as the percentage of correct selection of $\ubeta$ out of $R$ simulations. For example, if $\beta_d\neq 0$ and out of $R$ simulations, $\beta_d$ is selected to be non-zero for $g$ times, then the oracle percentage of $\beta_d$ is $\frac{g}{R}\times 100\%$.
MSE of $\beta_{d}$ is calculated as $\frac{1}{R}\sum_{ \ell = 1}^R(\hat{\beta}_{d}^{(\ell)} - \beta_{d})^2 $ where $\hat{\beta}_{d}^{(\ell)}$ is the estimator for $\beta_d$ in the $\ell$th simulation.

The simulation data were generated according to model (\ref{VMICM}),
where $\uX$ were generated from a $Unif(0,1)$ distribution. For the loading parameter $\ubeta = (\beta_1,\beta_2,\ldots,\beta_q)^T$, $\beta_{1} = \beta_2 = \frac{1}{\sqrt{2}}$ and the rest $\beta_{j}'s$ were set as zeros. We evaluated the performance of the proposed approach with both continuous and discrete predictors $G_{\cdot k}~(k=1,2,\ldots,p)$. For continuous variables $G_{\cdot k}~(k=1,2,\ldots,p)$, they can be gene expressions. For discrete variables $G_{\cdot k}~(k=1,2,\ldots,p)$, they can be single nucleotide polymorphism (SNP) variants. In either case, the dimension $p$ can be large.

\subsection{The Continuous Cases}
In the continuous case, the nonparametric functions $f_k(u)~(k=0,1,2,\ldots, p)$ were defined as follows: $f_0(u) = 2sin(2\pi u)$, $f_1(u) = 2cos(\pi u) + 2$ and $f_2(u) = sin(2\pi u) + cos(\pi u) + 1$ are varying functions; $f_3(u) = 2$ and $f_4(u) = 2.5$ are non-zero constants; $f_k(u) = 0$ are zeros for $k = 5,\ldots, p$. The number of loading parameters was set as $q = 5$ and $\beta_1 = \beta_2 = \frac{1}{\sqrt{2}}$, $\beta_3 = \beta_4=\beta_5 = 0$. Both $G_{\cdot k}~(k=1,2,\cdots, p)$ and $\uepsilon$ were generated from independent $ N(0,1)$. We run 1000 simulations (R = 1000) to evaluate the performance of the proposed variable selection approach under $p = 50, 100$.

Table \ref{cont_gamma} demonstrates the selection and estimation accuracy for continuous $G_{\cdot k}$. The left and right penal corresponds to the case where $p = 50$ and $100$ respectively.
For all the cases, the selection accuracy (oracle \%) is very closed to 100\% ($>99\%$), IMSE for varying functions ($f_0(\cdot), f_1(\cdot)$ and $f_2(\cdot)$) are in the order of $-2$, and IMSE for non-zero constant functions ($f_3(\cdot)$ and $f_4(\cdot)$) are in the order of $-3$. All of the model IMSE and oracle IMSE are in the same order. These observations indicate that our proposed estimation and selection approach possesses reasonable selection and estimation accuracy for the non-parametric function $f_k(\cdot)~(k=1,2,\ldots,p)$.

\begin{table}[H]\small
\tabcolsep 5pt
  \centering
  \caption{Selection (\%) and estimation accuracy (IMSE) of $f_k(\cdot)$ for continuous $G$.}
  \begin{tabular}{ccccccccc}
    \hline
       \multirow{2}{*}{Sample size}& \multirow{2}{*}{Function} & \multicolumn{3}{c}{$p = 50$} && \multicolumn{3}{c}{$p = 100$} \\
          \cline{3-5}\cline{7-9}
          &       & Oracle \% & Model & Oracle & &Oracle \% & Model & Oracle \\
    \hline
    \multirow{6}[2]{*}{$n = 500$} & $f_0(\cdot)$ & 100.0\% & 3.87E-02 & 4.27E-02 && 100.0\% & 3.77E-02 & 4.51E-02 \\
          & $f_1(\cdot)$ & 99.6\% & 1.58E-02 & 2.42E-02 && 99.9\% & 1.57E-02 & 3.14E-02 \\
          & $f_2(\cdot)$ & 99.9\% & 2.33E-02 & 2.58E-02 && 99.9\% & 2.26E-02 & 2.96E-02 \\
          & $f_3(\cdot)$ & 100.0\% & 2.09E-03 & 2.11E-03 && 100.0\% & 1.90E-03 & 1.97E-03 \\
          & $f_4(\cdot)$ & 100.0\% & 2.04E-03 & 2.06E-03 && 100.0\% & 2.07E-03 & 2.12E-03 \\
          & Zero  & 99.7\% & 1.94E-05 & 0 && 99.9\% & 1.12E-05 & 0 \\
    \hline
    \multirow{6}[2]{*}{$n = 1000$} & $f_0(.)$ & 100.0\% & 3.23E-02 & 3.40E-02 && 100.0\% & 3.31E-02 & 3.47E-02 \\
          & $f_1(\cdot)$ & 100.0\% & 7.17E-03 & 1.21E-02 && 100.0\% & 7.07E-03 & 1.17E-02 \\
          & $f_2(\cdot)$ & 100.0\% & 1.46E-02 & 1.59E-02 && 100.0\% & 1.46E-02 & 1.64E-02 \\
          & $f_3(\cdot)$ & 100.0\% & 1.02E-03 & 1.02E-03 && 100.0\% & 9.60E-04 & 9.55E-04 \\
          & $f_4(\cdot)$ & 100.0\% & 1.09E-03 & 1.09E-03 && 100.0\% & 1.06E-03 & 1.07E-03 \\
          & Zero  & 99.8\% & 8.50E-06 & 0 && 99.9\% & 3.46E-06 &  0 \\
    \hline
    \end{tabular}%
  \label{cont_gamma}%
\end{table}%

Table \ref{cont_beta} presents the selection and estimation accuracy for the loading parameter $\ubeta$. The results shows that the selection accuracy for all $\beta$ is reasonably good ($>98\%$) in all cases. For most of the $\beta$, the MSE is in the order of -4 or lower, except for $\beta_2$, which is -3 for both $p=50$ and $p=100$ when $n = 500$. The order of the model estimation for $\ubeta$ are at least the same as that of the oracle model if not lower. These results indicate that our model possesses good selection and estimation accuracy for the loading parameters $\ubeta$.

\begin{table}[H]\small
\tabcolsep 5pt
  \centering
  \caption{Selection (\%) and estimation accuracy (MSE) of $\ubeta$ for continuous $G$.}
    \begin{tabular}{ccccccccc}
    \hline
          \multirow{2}{*}{Sample size}& \multirow{2}{*}{$\beta$} & \multicolumn{3}{c}{$p = 50$} && \multicolumn{3}{c}{$p = 100$} \\
          \cline{3-5}\cline{7-9}
          &       & Oracle \% & Model & Oracle & &Oracle \% & Model & Oracle \\
 \hline
    \multirow{5}[2]{*}{$n = 500$} & $\beta_1$ & 100.0\% & 1.15E-04 & 1.07E-04 && 100.0\% & 1.17E-04 & 1.30E-04 \\
          & $\beta_2$ & 100.0\% & 8.04E-03 & 4.12E-03 && 100.0\% & 2.26E-03 & 7.62E-03 \\
          & $\beta_3$ & 98.1\% & 9.98E-05 & 0 && 98.2\% & 3.64E-05 & 0 \\
          & $\beta_4$ & 98.8\% & 2.99E-05 & 0 && 99.1\% & 3.13E-05 & 0 \\
          & $\beta_5$ & 98.6\% & 1.00E-04 & 0 && 98.5\% & 7.73E-05 & 0 \\
    \hline
    \multirow{5}[2]{*}{$n = 1000$} & $\beta_1$ & 100.0\% & 5.30E-05 & 5.52E-05 && 100.0\% & 5.00E-05 & 5.49E-05 \\
          & $\beta_2$ & 100.0\% & 5.34E-05 & 1.86E-03 && 100.0\% & 5.04E-05 & 1.79E-03 \\
          & $\beta_3$ & 98.9\% & 9.36E-06 & 0 && 98.8\% & 1.16E-05 & 0 \\
          & $\beta_4$ & 99.4\% & 6.30E-06 & 0 && 99.5\% & 5.49E-06 & 0 \\
          & $\beta_5$ & 99.1\% & 7.17E-06 & 0 && 99.0\% & 6.93E-06 & 0 \\
    \hline
    \end{tabular}
  \label{cont_beta}%
\end{table}%

\subsection{The Discrete Case}
We further evaluated how the proposed model performs with discrete $G_{\cdot k}~(k=1,2,\ldots,p)$, i.e., SNP data. In this simulation, each $G_{\cdot k}~(k=1,2,\ldots,p)$ variable was simulated from a multinomial distributions with minor allele frequency (MAF) $P_a$. The $G_{\cdot k}~(k=1,2,\ldots,p)$ variable takes values $0, 1, 2$ corresponding to the genotype $aa$, $Aa$, and $AA$ with corresponding genotype frequency $P_a^2$, $2P_a(1-P_a)$ and $(1-P_a)^2$, respectively. We set $P_a = 0.5$ for $k = 1,2,7$; $P_a = 0.3$ for $k = 3,4,8$; $P_a = 0.1$ for $k = 5,6,9$ and $P_a\sim Unif(0.05,0.5)$ for $k = 10,11,\ldots, p$.
For the non-parametric functions, $f_0(u) = 2sin(2\pi u)$, $f_1(u) = f_3(u) = f_5(u) = 2cos(\pi u) + 2$, $f_2(u) = f_4(u) = f_6(u) = sin(2\pi u) + cos(\pi u) + 1$; $f_7(u) = f_8(u) = f_9(u) = 2$; and $f_k(u) = 0$ for $k = 10,11,\ldots, p$.
Under the setup, we had both varying and constant effect with different minor allele frequencies.
$\uX$ was generated from $Unif(0,1)$ and $\epsilon$ was generated from $ N(0,1)$.
Finally, $Y$ was generated according to model (\ref{VMICM}). We evaluated the performance of the proposed model via $R=1000$ simulations under $p = 50, 100$ and $n = 500, 1000$.

Table \ref{discrete_gamma} presents the selection and estimation accuracy of the non-parametric function $f_k(\cdot)$. We observed that the oracle percentage are very high ($>98.8\%$) for all cases, indicating our proposed model can correctly select the coefficient functions with high accuracy. Further, the IMSE for varying functions are of the order $-2$ or lower, while the IMSE for constant functions are of the order $-3$ or lower. Moreover, the IMSE of the proposed model are in the same order of the IMSE of the oracle model. These suggest that our model performs reasonably well in both selection and estimation for the non-parametric functions.

\begin{table}[H]\small
\tabcolsep 5pt
  \centering
  \caption{Selection (\%) and estimation accuracy (IMSE) of $f_k(\cdot)$ for discrete $G$.}
    \begin{tabular}{ccccccccc}
    \hline
          \multirow{2}{*}{Sample size}& \multirow{2}{*}{Function} & \multicolumn{3}{c}{$p = 50$} && \multicolumn{3}{c}{$p = 100$} \\
          \cline{3-5}\cline{7-9}
          &       & Oracle \% & Model & Oracle & &Oracle \% & Model & Oracle \\
 \hline
    \multirow{11}[0]{*}{$n = 500$} & $f_0(.)$ & 100.0\% & 5.94E-02 & 5.66E-02 && 100.0\% & 7.42E-02 & 6.95E-02 \\

          & $f_1(\cdot)$ & 98.9\% & 3.71E-02 & 4.87E-02 && 98.4\% & 3.78E-02 & 5.44E-02 \\
          & $f_2(\cdot)$ & 99.0\% & 4.14E-02 & 3.79E-02 && 98.6\% & 4.30E-02 & 4.09E-02 \\
          & $f_3(\cdot)$ & 99.0\% & 3.50E-02 & 4.76E-02 && 98.5\% & 3.64E-02 & 5.81E-02 \\
          & $f_4(\cdot)$ & 98.9\% & 4.04E-02 & 3.63E-02 && 98.5\% & 4.48E-02 & 3.98E-02 \\
          & $f_5(\cdot)$ & 99.0\% & 4.02E-02 & 4.95E-02 && 98.6\% & 4.50E-02 & 7.29E-02 \\
          & $f_6(\cdot)$ & 98.8\% & 5.03E-02 & 4.52E-02 && 98.4\% & 4.98E-02 & 4.83E-02 \\
          & $f_7(\cdot)$ & 100.0\% & 2.37E-03 & 2.33E-03 && 99.9\% & 2.57E-03 & 2.51E-03 \\
          & $f_8(\cdot)$ & 100.0\% & 2.37E-03 & 2.37E-03 && 100.0\% & 2.55E-03 & 2.64E-03 \\
          & $f_9(\cdot)$ & 100.0\% & 2.66E-03 & 2.38E-03 && 100.0\% & 2.26E-03 & 2.24E-03 \\
          & Zero  & 99.6\% & 3.25E-05 & 0 && 99.7\% & 2.88E-05 & 0 \\
    \hline
    \multirow{11}[2]{*}{$n = 1000$} & $f_0(.)$ & 100.0\% & 3.12E-02 & 3.20E-02 && 100.0\% & 3.09E-02 & 3.44E-02 \\
          & $f_1(\cdot)$ & 99.9\% & 7.92E-03 & 1.22E-02 && 99.9\% & 7.96E-03 & 1.22E-02 \\
          & $f_2(\cdot)$ & 99.9\% & 1.50E-02 & 1.63E-02 && 99.9\% & 1.47E-02 & 1.59E-02 \\
          & $f_3(\cdot)$ & 99.9\% & 7.87E-03 & 1.21E-02 && 99.9\% & 8.19E-03 & 1.26E-02 \\
          & $f_4(\cdot)$ & 99.9\% & 1.44E-02 & 1.60E-02 && 99.9\% & 1.43E-02 & 1.58E-02 \\
          & $f_5(\cdot)$ & 99.9\% & 8.40E-03 & 1.17E-02 && 99.9\% & 8.54E-03 & 1.33E-02 \\
          & $f_6(\cdot)$ & 99.9\% & 1.48E-02 & 1.62E-02 && 99.9\% & 1.44E-02 & 1.64E-02 \\
          & $f_7(\cdot)$ & 100.0\% & 1.13E-03 & 1.14E-03 && 100.0\% & 9.55E-04 & 9.50E-04 \\
          & $f_8(\cdot)$ & 100.0\% & 1.14E-03 & 1.20E-03 && 100.0\% & 1.12E-03 & 1.16E-03 \\
          & $f_9(\cdot)$ & 100.0\% & 1.03E-03 & 1.04E-03 && 100.0\% & 1.13E-03 & 1.14E-03 \\
          & Zero  & 99.8\% & 9.21E-06 & 0 && 99.9\% & 4.38E-06 & 0 \\
    \hline
    \end{tabular}
  \label{discrete_gamma}%
\end{table}%

Table \ref{discrete_beta} presents the selection and estimation result of the loading parameters $\ubeta$. We observed that the oracle percentage in all the cases are above $98\%$, and the MSE for the estimation of $\ubeta$ is in the order of $-3$ or lower in the proposed and oracle model. These suggests that our proposed model can correctly select and estimate the loading parameters with high accuracy.

\begin{table}[H]\small
\tabcolsep 5pt
  \centering
  \caption{Selection (\%) and estimation accuracy (MSE) of $\ubeta$ for discrete $G$.}
    \begin{tabular}{ccccccccc}
    \hline
          \multirow{2}{*}{Sample size}& \multirow{2}{*}{$\beta$} & \multicolumn{3}{c}{$p = 50$} && \multicolumn{3}{c}{$p = 100$} \\
          \cline{3-5}\cline{7-9}
          &       & Oracle \% & Model & Oracle & &Oracle \% & Model & Oracle \\
 \hline
     \multirow{5}[2]{*}{n = 500} & $\beta_1$ & 100.0\% & 1.15E-04 & 1.07E-04 && 100.0\% & 1.17E-04 & 1.30E-04 \\
          & $\beta_2$ & 100.0\% & 8.04E-03 & 4.12E-03 && 100.0\% & 2.26E-03 & 7.62E-03 \\
          & $\beta_3$ & 98.1\% & 9.98E-05 & 0 && 98.2\% & 3.64E-05 & 0 \\
          & $\beta_4$ & 98.8\% & 2.99E-05 & 0 && 99.1\% & 3.13E-05 & 0 \\
          & $\beta_5$ & 98.6\% & 1.00E-04 & 0 && 98.5\% & 7.73E-05 & 0 \\
    \hline
    \multirow{5}[2]{*}{n = 1000} & $\beta_1$ & 100.0\% & 5.30E-05 & 5.52E-05 && 100.0\% & 5.00E-05 & 5.49E-05 \\
          & $\beta_2$ & 100.0\% & 5.34E-05 & 1.86E-03 && 100.0\% & 5.04E-05 & 1.79E-03 \\
          & $\beta_3$ & 98.9\% & 9.36E-06 & 0 && 98.8\% & 1.16E-05 & 0 \\
          & $\beta_4$ & 99.4\% & 6.30E-06 & 0 && 99.5\% & 5.49E-06 & 0 \\
          & $\beta_5$ & 99.1\% & 7.17E-06 & 0 && 99.0\% & 6.93E-06 & 0 \\
    \hline
    \end{tabular}
  \label{discrete_beta}%
\end{table}%

In all the simulation studies, we observed improved performance when the sample size increases from 500 to 1000. For example, as shown in Table \ref{discrete_beta}, the MSE for $\beta_5$ reduces from 1E-04 to 7.17E-06 when the sample size increases from 500 to 1000.

\section{Real Data Application}
We demonstrated the utility of the model with a human liver cohort (HLC) data set. The data set can be downloaded from www.synapse.org using synapse ID: syn4499 which contains gene expressions and phenotypes (activity of several liver enzymes). For more details regarding the data set, please refer to Schadt et al. \cite{Schadt2008} and Yang et al. \cite{Yang2010}. In the HLC data set, the phenotypes are enzyme activity measurements of Cytochrom P450. There are a total of nine P450 enzymes (CYP1A2, 2A6, 2B6, 2C8, 2C9, 2C19, 2D6, 2E1, and 3A4). We chose CYP2E1 to demonstrate the utility of the method. 
For the environmental variable ($\uX$), we chose Age (=$X_1$), Aldehyde Oxydase ($X_2$), and Liver Triglyceride ($X_3$), then transformed each one of them to [0,1] with $\frac{X_{i} - \min(X_{i})}{\max(X_{i}) - \min(X_{i})}$. In this analysis, we focused on gene expressions which are treated as the $G$ variable. After data cleaning, we had $n = 394$ (sample size) and $N = 19,172$ (number of gene expressions). Applying the proposed method, we would like to answer the following questions: (1) which gene is sensitive to the synergistic effect of the three $X$ variables to affect the CYP2E1 activity? (2) what is the effect function of the three $X$ variables as a whole, zero, constant or varying? and (3) which $X$ variable contributes to the synergistic interaction effect?

We focused on the KEGG pathway ``Metabolism of Xenobiotics by Cytochrome P450" (hsa00980) to select important genes associated with CYP2E1 activity.
There are 76 genes in this pathway and 70 are mapped to our data set.
After applying the proposed method, we identified one gene expression (SULT2A1) with varying effect and three gene expressions (FABP1, C15orf39, B3GNT5) with constant effect.

\begin{figure}[H]
\small
\centering
\includegraphics[width=10cm,height=8cm]{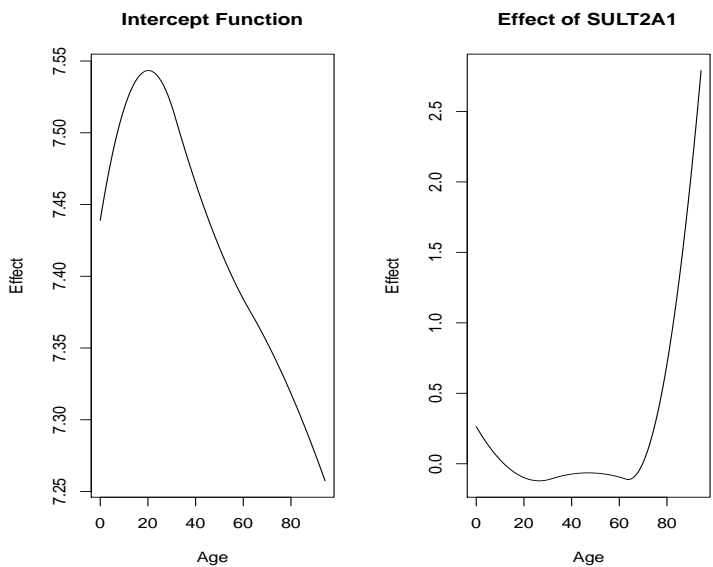}
\label{realdata1}
\caption{Plot of the varying effect for gene SULT2A1.}\label{fig1}
\end{figure}

Figure \ref{fig1} presents the plot of the intercept function (left panel) and the varying coefficient function for gene \emph{SULT2A1} (right panel) on CYP2E1 activity. After shrinkage, the coefficients for $X_2$ and $X_3$ were all zeros, leaving only Age as the effective environmental factor. The intercept function first increases before age 20, then it decreases dramatically for the rest of the life, showing the overall declining CYP2E1 enzyme activity over age. The effect of gene SULT2A1 on the CYP2E1 activity, however, behaves quite differently. The effect of this gene on CYP2E1 activity shows little change (around the zero line) before age 65. After that, it shows a positive effect on CYP2E1 activity as people become old. Gene \emph{SULT2A1} encodes sulfotransferase which aids in the metabolism of drugs and endogenous compounds. Study by Echchgadda et al. \cite{Echchgadda2004} showed that in senescent male rodents, \emph{Sult2A1} gene transcription in the liver is significantly enhanced due to the age-associated loss of the liver expression of androgen receptor. Although the study was conducted in rodents, it has implication on humans. Our result of enhanced function of \emph{SULT2A1} late in life agrees with the finding by Echchgadda et al. \cite{Echchgadda2004}.
This result also demonstrates the unique strength of the proposed method to capture the non-linear interaction between environmental factors and genes. However, further biological investigation is needed to confirm the real function of this gene modified by aging. In addition to this gene, genes with constant effect are FABP1 ($\hat{f}=0.135$), C15orf39 ($\hat{f}=-0.112$) and B3GNT5 ($\hat{f}=-0.128$). The constant effects indicate that the effect of these genes on CYP2E1 does not change over age. In addition, the negative effect size tells that the CYP2E1 activity is negatively regulated by these genes. We did not find literature report to support that these genes show age-related expressions.

\section{Discussion}
VMICM is a promising tool to model non-linear interactions between genes and multiple environments as a whole.
It combines multiple exposure variables $\uX$ into a single-index $\uX^T\ubeta$, hence can reduce model dimension and alleviate the curse of dimensionality.
In this paper, we develop a three stage variable selection approach for VMICM.
Our goal is to identify varying, non-zero constant and zero effects which respectively correspond to nonlinear G$\times$E effect, no G$\times$E effect and no genetic effect.
In the meantime, we also select important exposure variables.
Rather than modeling the G$\times$E effect for each $X$ variable separately, our approach can model the joint effect of multiple environmental factors ($\uX$) as a whole, then identify how different genes interact with the environmental mixture to affect a disease trait, the so called synergistic G$\times$E interaction. Our model is biologically motivated and attractive since it offers an alternative strategy to look for G$\times$E interaction. In addition, our model is flexible to detect any potential non-linear interactions. We further studied the theoretical property of the proposed estimation and selection method. Both simulation and real data analysis demonstrate the utility of the proposed method.

In our model setup, the covariates $\uX$ are assumed to be continuous. This is due to the fact that the index $u=\uX^T\ubeta$ has to be continuous in order to model the nonlinear function.
In real applications, environmental variables can be discrete such as smoking, gender and ethnicity group. To accommodate the presence of discrete factors, the VMICM can be generalized to a partial linear VMICM, i.e.,
\begin{equation}\label{PLVMICM}
  \uY = \uf(\uX^T\ubeta)\uG +\uZ\ualpha + \uZ\uG\udelta + \uepsilon
\end{equation}
where $\uZ $ represent discrete covariates and $\ualpha$ and $\udelta$ represent the effects of $\uZ$ and the interaction between $\uZ$ and $\uG$, respectively.
According to (\ref{Bsplpprox})-(\ref{VMICMapprox2}), we have
\begin{equation}\label{PLVMICMapprox}
    \uY \approx \uW(\ubeta)\ugamma + \uZ\ualpha + \uZ\uG\udelta + \uepsilon
\end{equation}
Our variable selection approach could be modified slightly to perform selection of non-parametric functions and the parametric components simultaneously. More specifically, the design matrix can be updated to $(\uW(\ubeta),\uZ, \uZ\uG)$ in Step 1 in the algorithm, then the rest follows.

So far we discussed the variable selection approach for VMICM with a continuous response phenotype. In practice, many phenotype can be categorical such as a binary disease response in a case control study. It is natural to extend the current selection approach to a generalized VMICM framework, which will be investigated in our future work.

In our model formulation, we assumed different index coefficients share common loading parameters, i.e., $\ubeta_0=\ubeta_1=\cdots=\ubeta_p=\ubeta$. From a practical point of view, assuming different loading parameters makes perfect sense such as the model proposed by Ma and Song \cite{MA2015a}. However, such a treatment imposes theoretical challenges when evaluating the theoretical properties such as the selection consistency. This is because that the loading coefficients for the $k$th index coefficient are not identifiable when $f_k(u)\notin \mathcal{V}$. When a coefficient function is not varying, $\ubeta_k$ does not exists. Thus, the selection consistency for $\ubeta_k$ does not exists. For this reason, we impose the same loading parameters for all the index coefficient functions.
In addition to the application to G$\times$E studies, our model has many applications in other fields where the purpose is to model the interaction between one variable and a mixture of a few other variables, the so called synergistic interaction.


\section*{Appendix}\label{appendix}
\setcounter{equation}{0}
\renewcommand{\theequation}{A-\arabic{equation}}
\renewcommand{\thesubsection}{A.\arabic{subsection}}
\subsection{Computational Algorithms}

From (\ref{VMICMapprox2}), we have the design matrix $\uW(\ubeta)$
with the corresponding parameters $\ugamma = (\ugamma_0^T,\ugamma_1^T, \ldots,\ugamma_p^T)^T$ and $\ugamma_k=(\gamma_{k1}, \ugamma_{k*}^T)^T$.
Then the detailed computational algorithms for Step 1, Step 2 and Step 3 are given as follows.

\textbf{Computational algorithm for Step 1}:
In this step, we get the estimator $\hat\ugamma^{(1)}$ denoted in (\ref{gammahat(1)}) by minimizing the objective function $Q_1(\ugamma|\Lambda_1,\hat{\ubeta}^{(0)})$ and using the group coordinate descent algorithm for iterative computation.
We first assign a grouping index from $0$ to $M$ for each of the parameters.
Furthermore, parameters with the same grouping index are in the same group and penalized as a group.
Parameters with grouping index 0 are not penalized.
Clearly, $\{\ugamma_k,k=0,1,\dots,p\}=\{\ugamma_{(m)},m=0,1,\dots,M\}$, and $\{\hat\ugamma_k,k=0,1,\dots,p\}=\{\hat\ugamma_{(m)},m=0,1,\dots,M\}$.
Denote $\uW_{(m)}$ as the design matrix for group $m$, $ m = 0,1,\ldots M$. Given a tuning parameter $\lambda$ and MCP tuning parameter $\tau^{MCP}$, $\hat \ugamma^{(1)}$ can be obtained through the following iteration.
\begin{itemize}
\item[(0)] Run a Q-R decomposition on all $\uW_{(m)}$, i.e., $\uW_{(m)} = \uQ_{(m)} \uR_{(m)}$, $m = 0,1,2\ldots M$, where $\uQ_{(m)}^T\uQ_{(m)} = \uI$ and $\uR_{(m)}$ is an upper triangular matrix, $\uQ_{(m)}$ is the normalized design matrix for group $m$.

\item[(1)] Assign the grouping index for the initial values $\hat\ugamma^{(0)}$ from (\ref{gammahat0}) such as $\{\hat{\ugamma}_{(m)}^{(0)}, m = 0,1,\ldots ,M\}$, obtain the ordinary least squares (OLS) estimator $\hat{\ugamma}_{(m)}^{OLS}$ via $\hat{\ugamma}_{(m)}^{OLS} = \uQ_{(m)}^T(\uY - \uQ_{-(m)}\hat{\ugamma}_{-(m)}) = \uQ_{(m)}^T\uY -  \uQ_{(m)}^T\uQ_{-(m)}\hat{\ugamma}_{-(m)}$, where subscript $\uQ_{-(m)}$ represents the normalized design matrix without group $m$ and $\hat{\ugamma}_{-(m)}$ represents the most updated values for $\ugamma$ without group $m$.

\item[(2)] For $m=0$, set $\hat{\ugamma}_{(0)} = \hat{\ugamma}^{OLS}_{(0)}$.

\item[(3)] For $m = 1,\ldots, M$, obtain the MCP estimate $\hat{\ugamma}_{(m)}$ via

\begin{equation}\label{A1}
  {\hat \ugamma_{(m)}} = \left\{ {\begin{array}{*{20}{l}}
{\hat \ugamma_{(m)}^{OLS},}&\text{if}~~{\| {\hat \ugamma_{(m)}^{OLS}} \| > \lambda \tau^{MCP} }\\
{\frac{\tau }{{1 - \tau }}S(\hat \ugamma_{(m)}^{OLS},\lambda ),}&\text{if}~~{\| {\hat \ugamma_{(m)}^{OLS}} \| \le \lambda \tau^{MCP} }
\end{array}} \right. , ~~~~m=1,2,\ldots,M
\end{equation}
where $S(\hat{\ugamma}_{(m)}^{OLS},\lambda) = \hat{\ugamma}_{(m)}^{OLS}\left(1 - \frac{\lambda}{\|\hat{\ugamma}_{(m)}^{OLS}\|}\right)_{+} $.
%

\item[(4)] Updated $\hat{\ugamma}_{(m)}^{(0)}$ in step (1) by $\hat{\ugamma}_{(m)}$.
\end{itemize}

Iterate step (1) through step (4) until convergence and get an unadjusted MCP estimator denoted as $\hat{\ugamma}^{unadjusted}$.
Then, we can get an adjusted MCP estimator as
\begin{equation}\label{gammahatm}
  \hat{\ugamma}_{(m)} = \uR_{(m)}^{-1}\hat{\ugamma}_{(m)}^{unadjusted},~~~~m=0,1,\ldots, M
\end{equation}
Accordingly, we have $\{\hat\ugamma_k^{(1)},k=0,1,\dots,p\}=\{\hat\ugamma_{(m)},m=0,1,\dots,M\}$.
Finally, we can get our Step 1 estimator $\hat{\ugamma}^{(1)}=((\hat\ugamma_0^{(1)})^T, (\hat\ugamma_1^{(1)})^T,\ldots, (\hat\ugamma_p^{(1)})^T)^T$.

\textbf{Computational algorithm for Step 2}:
In this step, given $\hat{\ugamma}^{(1)}$ in Step 1, we get the estimator $\hat\ugamma^{(2)}$ denoted in (\ref{gammahat(2)}) and use the group coordinate descent algorithm for iterative computation, same as in Step 1.
We first get different design matrix and grouping index according to $\hat{\ugamma}^{(1)}$;
then, repeat Step 1 until convergence to get $\hat{\ugamma}^{(2)}$.

\textbf{Computational algorithm for Step 3}:
In this step, given $\hat{\ugamma}^{(2)}$ in Step 2, we get $\hat\ubeta$ denoted in (\ref{betahat}).
We adopt the idea of first order approximation and coordinate decent algorithm to estimate $\ubeta$ by minimizing (\ref{Q3}).
Since $\bar{\uB}(\uX^T\ubeta)$ is not a linear function of $\ubeta$, there is no closed form solution of $\ubeta$. Hence, we apply a local linear approximation of $\bar{\uB}(\uX^T\ubeta)$ at $\tilde{\ubeta}$, and $\tilde{\ubeta}$ is the most updated value of $\ubeta$. We have
\begin{equation}
  \bar{B}(\uX^T\ubeta) \hat{\ugamma}_{k*} \approx \bar{B}(\uX^T\tilde{\ubeta}) \hat{\ugamma}_{k*} + \bar{B}'(\uX^T\tilde{\ubeta})  \hat{\ugamma}_{k*}\uX(\ubeta - \tilde{\ubeta} )
\end{equation}

Working with $\beta_d$, $d=1,2,\ldots,q$, we have
\begin{equation}\label{S4}
  \bar{B}(\uX^T\ubeta) \hat{\ugamma}_k^* \approx \bar{B}(\uX^T\tilde{\ubeta}) \hat{\ugamma}_k^*  + \bar{B}'(\uX^T\tilde{\ubeta}) \hat{\ugamma}_k^* \uX_d (\beta_{d} - \tilde{\beta}_{d})
\end{equation}

Then we can obtain $\hat{\beta}_{d}$ by minimizing the following penalized function,
\begin{equation}
  Q_{d} = \| \uY_{d}^* - \uX_{d}^*\beta_{d} \|^2 + n p_{\lambda_3}(|\beta_{d}|)
\end{equation}
where $$ \uY_{d}^* = \uY - \sum_{k=0}^p[ \hat{\gamma}_{k1}\uG_k + \bar{B}(\uX^T\tilde{\ubeta}) \hat{\ugamma}_k^* \uG_k - \bar{B}^T(\uX\tilde{\ubeta}) \hat{\ugamma}_k^* \uG_k \uX_d \tilde{\beta}_{d}],$$
$$\uX_{d}^* = \sum_{k=0}^p \bar{B}^T(\uX\tilde{\ubeta}) \hat{\ugamma}_k^* \uG_k \uX_d.$$
Then, the MCP penalized estimator $\hat{\ubeta}^* = (\hat\beta_1^*,\ldots,\hat\beta_q^*)^T$ can be obtained via the coordinate descent algorithm.
Since there are two constrains on $\ubeta$: (1) $\|\ubeta\|_2 = 1$ and (2) $\beta_{1} > 0$. We do not penalize $\beta_{1}$ and normalize $\ubeta$ after updating $\ubeta$, i.e., $\hat{\beta}_{d} = \frac{\hat{\beta}_{d}^*}{\|\hat{\ubeta}^*\|}\sgn(\hat{\beta}_{1}^*)$.
The detailed algorithm for estimating $\beta_d, d=1,2,\ldots,q$, is given as follows:

(0) Get the initial estimator $\hat\ubeta^{(0)}$ from (\ref{betahat0});

(1) Calculate $\uY_{d}^*$ and $\uX_{d}^*$;

(2) Normalized $\uX_{d}^*$ by $ \tilde\uX_{d}^* =  \uX_{d}^*/\|\uX_{d}^*\|$;

(3) Calculate $\hat{\beta}_{d}^{OLS} = \tilde\uX_{d}^{*T}\uY_{d}^*$

(4) Let $\hat{\beta}_{1}^* = \hat{\beta}_{1}^{OLS}$ and for $d\neq 1, \hat{\beta}_{d}^* = \frac{(\hat{\beta}_{d}^{OLS} - \lambda)_+}{1-1/\tau^{MCP} }$ if $|\hat{\beta}_{d}^{OLS}|\leq\lambda\tau^{MCP}$ and $\hat{\beta}_{d}^* = \hat{\beta}_{d}^{OLS}$ if $|\hat{\beta}_{d}^{OLS}|>\lambda\tau^{MCP}$;

(5) Normalized $\hat{\ubeta}^* = (\hat{\beta}_1^*,\ldots,\hat{\beta}_q^*)^T$, i.e., $\hat{\beta}_{d} = \frac{\hat{\beta}_{d}^*}{\|\hat{\ubeta}^*\|}\sgn(\hat\beta_{1}^*)$;

(6) Update $\hat\ubeta^{(0)}$ in step (0) with $\hat{\ubeta} = (\hat{\beta_1},\ldots,\hat{\beta_q})^T$, then iterate until convergence.

\subsection{Proofs of Theorems}


The following regularity conditions are assumed.

(A1) The density function $f_U(u)$ of a random variable $U = \uX^T\ubeta$ is bounded away from $0$ on $\mathcal{U} = \{ u = \uX^T\ubeta: \uX\in \mathcal{X} \}$, where $\mathcal{X} $ is the compact support of $\uX$. Furthermore, we assume that $f_u(\cdot)$ satisfies the Lipschitz condition of order 1 on $\mathcal{U}$;

(A2) $f_k(\cdot)~(k = 0,1,\ldots, p )$ have bounded and continuous derivatives up to order $r$ on $\mathcal{U} $ and $r\geq 2$;

(A3) $\E(\|\uG\|^6) < \infty$ and $\E(|\epsilon|^6) < \infty$;

(A4) $\{(Y_i, \uX_i, \uG_i), i=1,2,\ldots,n\}$ is a strictly stationary and strongly mixing sequence with mixing coefficient $\alpha(n) = O(\rho^n)$ for some $0<\rho<1$;

(A5) Let $b_n = \max_{k,l}\{ p''_{\lambda_1}(\|\ugamma_{k*}^0\|), p''_{\lambda_2}(|\ugamma_{k1}^0|), p''_{\lambda_3}(|\beta_{d}^0|) , \ugamma_{k*}^0\neq 0,\ugamma_{k1}^0\neq 0,\beta_l^0\neq 0  \}$
for $k = 1,\ldots, p, d = 2,\ldots, q$, then $b_n\to 0$ as $n\to 0$;

(A6) $\liminf_{n\to\infty}\liminf_{\|\ugamma_{k*}\|\to 0^+} \frac{1}{\lambda_1}| p'_{\lambda_1}(\|\ugamma_{k*}\|)| > 0$ for $k = v+1, \ldots, p
$
$$
\liminf_{n\to\infty}\liminf_{|\ugamma_{k1}|\to 0^+} \frac{1}{\lambda_2}| p'_{\lambda_2}(|\ugamma_{k1}|)| > 0\ {\rm for} \ k = c+1, \ldots, p
$$
\[
\liminf_{n\to\infty}\liminf_{|\beta_{d}|\to 0^+} \frac{1}{\lambda_3}| p'_{\lambda_3}(|\beta_{d}|)| > 0\ {\rm for}\ d = s+1, \ldots, q
\]

(A7) Let $\kappa_1, \kappa_2,\ldots, \kappa_K$ be internal knots of $[a,b]$, where $a = \inf\{ u: u\in \mathcal{U} \}$, $b = \sup\{ u: u\in \mathcal{U} \}$. Furthermore, let $\kappa_1 = a $, $\kappa_{K+1} = b$, $h_i = \kappa_i - \kappa_{i-1}$, $h_{\max} = \max\{ h_i\}$, $h_{\min} = \min\{ h_i\}$. Then, there exist a constant $C_0$ such that $\frac{h_{\max}}{h_{\min}} < C_0$ and $\max\{ h_{i+1} - h_i \} = o(K^{-1})$;

(A8) $D(u)$ is positive, and each element of $C_1(u)$ and $C_2(u)$ satisfy the Lipschitz condition of order 1 on $\mathcal{U}$.

Before the proof, we first define some notations as follows:
\begin{align*}
\uPsi_{11}& = \frac{1}{n}\sum_{i=1}^{n}\uW_{(\mathcal{V})i}^*(\uphi_0^*)\uW_{(\mathcal{V})i}^{*T}(\uphi_0^*),
\uPsi_{12} = \frac{1}{n}\sum_{i=1}^{n}\uW_{(\mathcal{V})i}^*(\uphi_0^*) \uW_{(\mathcal{C})i}^{*T}(\uphi_0^*),\\
\uPsi_{13} &= \frac{1}{n}\sum_{i=1}^{n}\uW_{(\mathcal{V})i}^*(\uphi_0^*)\uV_i^{*T},
\uPsi_{21} = \frac{1}{n}\sum_{i=1}^{n}\uW_{(\mathcal{C})i}^*(\uphi_0^*) \uW_{(\mathcal{V})i}^{*T}(\uphi_0^*),\\
\uPsi_{22} &= \frac{1}{n}\sum_{i=1}^{n}\uW_{(\mathcal{C})i}^*(\uphi_0^*) \uW_{(\mathcal{C})i}^{*T}(\uphi_0^*),
\uPsi_{23} = \frac{1}{n}\sum_{i=1}^{n}\uW_{(\mathcal{C})i}^*(\uphi_0^*)\uV_i^{*T}\\
\uPsi_{31} &= \frac{1}{n}\sum_{i=1}^{n}\uV_i\uW_{(\mathcal{V})i}^{*T}(\uphi_0^*),
\uPsi_{32} = \frac{1}{n}\sum_{i=1}^{n}\uV_i^*\uW_{(\mathcal{C})i}^*(\uphi_0^*)^{*T},\\
\uPsi_{33} &= \frac{1}{n}\sum_{i=1}^{n}\uV_i^*\uV_i^{*T},
\uLambda_{10} = \frac{1}{n}\sum_{i=1}^{n}\uW_{(\mathcal{V})i}^*(\uphi_0^*)(\epsilon_i + \uR^T(\uphi_0^*)\uG_i^*)
\end{align*}

\textbf{Lemma 1} If $f_k(u)~(k = 0,1,\ldots,p)$ satisfies condition (A2), then there exists a constat $C_0>0$ such that
\begin{equation}\label{Lemma1}
  \sup_{u\in\mathcal{U}}| f_k(u) - \uB^T(u)\ugamma_{k*}| \leq C_0 K^{-r}
\end{equation}

\textbf{Proof}: This result follows directly from the standard B-spline theory.

\textbf{Lemma 2}
Suppose the regularity conditions (A1) - (A7) hold and the number of knots $K = O_p(n^{1/(2r+1)})$. Then we have
\begin{equation}\label{Lemma2}
  \uPsi_{22}-\uPsi_{12}^T\uPsi_{11}^{-1}\uPsi_{12} \xrightarrow{P}\Sigma_1 ~{\rm and}~
  \uPsi_{33}-\uPsi_{13}^T\uPsi_{11}^{-1}\uPsi_{13} \xrightarrow{P}\Sigma_2
\end{equation}
where notation ``$\xrightarrow{P}$" represents convergence in probability.

\textbf{Proof}: The results of this lemma follow directly from \cite{Feng2013} and \cite{Zhao2010}.

\textbf{Proof of Theorem 1}:
To show the consistency of $\hat{\ubeta}$ is equivalent to show the consistency of $\hat{\uphi}$.
Let $\alpha_n = n^{-r/(2r+1)} + a_n$, $\uphi = \uphi^0 + \delta\utau_1$, $\ugamma = \ugamma^0 + \delta\utau_2$ and $\utau = (\utau_1^T, \utau_2^T)^T$,
where $ \utau_2 = (\tau_{01}, \utau_{0*},\ldots,\tau_{p1}, \utau_{p*}  ) $ and $\{\tau_{k1}, \utau_{k*}\}$ corresponds to the B-spline coefficients $\{\gamma_{k1},\ugamma_{k*}\}$;
$\utau_1 = (\tau_1^{\phi},\ldots,\tau_{q-1}^{\phi})$; $\tau_l^{\phi}$ corresponds to $\phi_l$;
and $\ugamma^0$ and $\uphi^0$ are the true value of $\ugamma$ and $\uphi$, respectively.

To show the consistency of $\hat{\ugamma}$ and $\hat{\uphi}$, we need to show $\forall \epsilon > 0$, $\exists$ a large enough $C$ such that
\begin{equation}\label{converge_prob}
  P\left\{\inf_{\|\utau\| = C} \{Q(\uphi,\ugamma)\} > Q(\uphi^0,\ugamma^0)\right\}\geq 1-\epsilon.
\end{equation}

If (\ref{converge_prob}) holds, we can say with probability at least $1 - \epsilon$, there exists a local minimum in the ball $\{ (\ugamma^0,\uphi^0) + \delta\utau:\|\utau\|\leq C \}$.
Hence, there exists a local minimizer such that $ \| (\hat{\ugamma},\hat{\uphi})-(\ugamma^0,\uphi^0) \| = O_p(\delta)$.

Let $D_n(\utau) = K^{-1}\{Q(\ugamma,\uphi) - Q(\ugamma^0,\uphi^0)\}$, we can get
\begin{align*}
D_n(\utau) &=  \frac{1}{K}\sum_{i=1}^{n}\left[\left( Y_i - \uW_i^T(\uphi^0+\delta \utau_1)(\ugamma^0+\delta \utau_2)\right)^2 - \left( Y_i - \uW_i^T(\uphi^0)\ugamma^0\right)^2\right] \\
&+ \frac{n}{K}\sum_{k=1}^p \left[ p_{\lambda_{1k}}(\|\ugamma_{k*}^0 + \delta\utau_{k*}\|) - p_{\lambda_{1k}}(\|\ugamma_{k*}^0 \|)\right]
&\\
&+  \frac{n}{K}\sum_{k=1}^p\left[p_{\lambda_{2k}}(|\gamma_{k1}^0 + \delta\tau_{k1}|)I(\|\ugamma_{k*}^0 + \delta\tau_{k*}\| = 0) -  p_{\lambda_{2k}}(|\gamma_{k1}^0|)I(\|\ugamma_{k*}^0\| = 0) \right]
&\\
&+ \frac{n}{K}\sum_{d=1}^{q-1} \left[p_{\lambda_{3d}}(|\phi_d^0 + \delta\tau_d^{\phi}|) - p_{\lambda_{3d}}(|\phi_d^0 |)\right]
\end{align*}

Since $p_{\lambda_{1k}}(\|\ugamma_{k*}^0 \|)] = 0$ for $k = v+1,\ldots,p$ and $p_{\lambda_{3d}}(|\phi_d^0 |) = 0$ for $d=s+1,\ldots,q-1$ and $I(\|\ugamma_{k*}^0\| = 0) = 0$ for $k = 1,\ldots,v$ , we have
\begin{align*}
D_n(\utau) &\geq  \frac{1}{K}\sum_{i=1}^{n}\left[\left( Y_i - \uW_i^T(\uphi^0+\delta \utau_1)(\ugamma^0+\delta \utau_2)\right)^2 - \left( Y_i - \uW_i^T(\uphi^0)\ugamma^0\right)^2\right] \\
&+ \frac{n}{K}\sum_{k=1}^v \left[ p_{\lambda_{1k}}(\|\ugamma_{k*}^0 + \alpha_n\utau_{k*}\|) - p_{\lambda_{1k}}(\|\ugamma_{k*}^0 \|)\right]
&\\
&+  \frac{n}{K}\sum_{k=v+1}^p\left[p_{\lambda_{2k}}(|\gamma_{k1}^0 + \delta\tau_{k1}|) -  p_{\lambda_{2k}}(|\gamma_{k1}^0|)\right]
&\\
&+ \frac{n}{K}\sum_{j=1}^{s-1} [p_{\lambda_{3d}}(|\phi_j^0 + \alpha_n\tau_j^{\phi}|) - p_{\lambda_{3d}}(|\phi_j^0 |)]
\end{align*}

By Taylor Expansion at $(\ugamma^0,\uphi^0)$, following \cite{Feng2013}, we have
\begin{align*}
D_n(\utau)&\geq  \frac{-2\delta}{K}\sum_{i=1}^{n}\left[ (\epsilon_i + R^T(\uX_i^T\ubeta^0)\uG_i)(\dot{\uW}_i^T(\uphi^0)\ugamma^0\uJ^T_{\uphi^0}\uX_i\utau_1 + \uW_i^T(\uphi^0)\utau_2)\right]\\
&+ \frac{\delta^2}{K}\sum_{i=1}^{n}(\dot{\uW}_i^T(\uphi^0)\ugamma^0\uJ^T_{\uphi^0}\uX_i\utau_1 + \uW_i^T(\uphi^0)\utau_2)^2+o_p(1)
\\
&+ \frac{n}{K}\sum_{k=1}^v\big[\delta p'_{\lambda_{1k}}(\|\ugamma_{k*}^0\|)\frac{\ugamma_{k*}^0}{\|\ugamma_{k*}^0\|}\utau_{k*}^T + \delta^2p''_{\lambda_{1k}}(\|\ugamma_{k*}^0\|)\utau_{k*}\utau_{k*}^T(1+o_p(1))\big]
&\\
&+ \frac{n}{K}\sum_{k=v+1}^p\big[ \delta p'_{\lambda_{2k}}(|\gamma_{k1}^0|)\sgn(\gamma_{k1}^0)\tau_{k1} + \delta^2p''_{\lambda_{2k}}(|\gamma_{k1}^0|)(\tau_{k1})^2(1+o_p(1))   \big ]
&\\
&+
\frac{n}{K}\sum_{d=1}^{s-1}\big[\delta p'_{\lambda_{3d}}(|\phi_d^0|)\sgn(\phi_d^0)\tau_d^{\phi} + \delta^2p''_{\lambda_{3d}}(|\phi_d^0|)(\tau_d^{\phi})^2(1+o_p(1))\big]
&\\
&=: S_1 + S_2 + o_p(1) + S_3 + S_4 + S_5
\end{align*}
where $\dot{\uW}_i(\uphi^0)= I_{p+1}\otimes \dot{\uB}(\uX_i^T\ubeta^0)\cdot \uG_i $,
$R(u)= (R_0(u),R_2(u),\ldots, R_p(u))^T$, $R_k(u)=f_k(u) - \uB^T(u)\ugamma_k^0$, $k=0, 1,\ldots, p$.
From Lemma 1, we have $|R_k(u)| = O(K^{-r})$ and
\begin{equation}\label{dotfk}
|\dot{f}_k(\uX_i^T\ubeta^0) - \dot{\uB}^T(u)\ugamma_k^0| \leq C_0 K^{-r+1}
\end{equation}

Note that $\epsilon_i$ is independent of $(\uX_i, \uG_i)$, we have
\begin{equation}\label{S11}
  \frac{1}{\sqrt{n}}\sum_{1}^{n}\epsilon_i (\dot{\uW}_i^T(\uphi^0)\ugamma^0\uJ^T_{\uphi^0}\uX_i\utau_1 + \uW_i^T(\uphi^0)\utau_2) = O_p(\|\utau\|)
\end{equation}
In addition, from (\ref{S4}), we can get
\begin{align}\label{S12}
  \sum_{i=1}^{n}& R^T(\uX_i^T\ubeta^0)\uG_i(\dot{\uW}_i^T(\uphi^0)\ugamma^0\uJ^T_{\uphi^0}\uX_i\utau_1 + \uW_i^T(\uphi^0)\utau_2)\nonumber\\
  &=\sum_{i=1}^{n}R^T(\uX_i^T\ubeta^0)\uG_i\{\dot{\uf}^T(\uX_i^T\ubeta^0)\uG_i\uJ^T_{\uphi^0}\uX_i\utau_1 \nonumber\\
  &+(\dot{\uW}_i^T(\uphi^0)\ugamma^0 - \dot{\uf}^T(\uX_i^T\ubeta^0)\uG_i\uJ^T_{\uphi^0}\uX_i\utau_1 + \uW_i^T(\uphi^0)\utau_2)\}\nonumber\\
  &= O_p(nK^{-r}\|\utau\|).
\end{align}
Following \cite{Feng2013}, from (\ref{S11}), (\ref{S12}) and (\ref{S4}), it is easy to show that
\begin{equation}\label{S1}
  S_1 = O_p(\sqrt{n}K^{-1}\delta)\|\utau\| + O_p(nK^{-1-r}\delta)\|\utau\| = O_p(1 + n^{r/(2r+1)}a_n)\|\utau\|.
\end{equation}
Similarly, we can get
\begin{equation}\label{S2}
  S_2 = O_p(\sqrt{n}K^{-1}\delta^2)\|\utau\|^2=O_p(1 + 2n^{r/(2r+1)}a_n)\|\utau\|^2.
\end{equation}
Hence, $S_2$ dominates $S_1$ uniformly in $\{\utau: \|\utau\| = C\}$ by choosing a sufficiently large $C$.

Further, by Taylor expansion at $\ugamma^0$, we have
\begin{align*}
S_3 &\leq nK^{-1}\delta a_n\sum_{k=1}^v\frac{\ugamma_{k*}^0}{\|\ugamma_{k*}^0\|}\utau_{k*}^T + nK^{-1}\delta^2 b_n \sum_{k=1}^v\utau_{k*}\utau_{k*}^T\\
&\leq nK^{-1}\delta a_n\sqrt{v}\|\utau\| + nK^{-1}\delta^2 b_n \|\utau\|^2
\end{align*}
Since $b_n\to 0$, then it is easy to show that $S_3$ is dominated by $S_2$ uniformly in $\|\utau\| = C$.

For $S_4$ and $S_5$, we have
\begin{align*}
S_4 & \leq  \delta a_n n K^{-1}\sum_{k=v+1}^p\tau_{k1} +  n K^{-1} \delta^2 b_n\sum_{k=v+1}^p(\tau_{k1})^2
 \leq  n K^{-1} \delta^2 C +  n K^{-1} \delta^2 C^2 b_n, \\
S_5 & \leq  \delta a_n n K^{-1}\sum_{j=1}^s\tau_j^{\phi} + n K^{-1} \delta^2 b_n\sum_{j=1}^s(\tau_j^{\phi})^2
 \leq n K^{-1}\delta^2 C + n K^{-1}\delta^2 C^2 b_n.
\end{align*}

With the same argument, we have $S_4$ and $S_5$ dominated by $S_2$ uniformly in $\|\utau\| = C$.
Hence, by choosing a large enough $C$, (\ref{converge_prob}) holds.
Therefore, there exists local minimizers $\hat{\uphi}$ and $\hat\ugamma$ such that
$$ \| \hat{\uphi}-\uphi^0 \| = O_p(\delta), ~\| \hat{\ugamma}-\ugamma^0 \| = O_p(\delta).$$
So we can get $\|\hat{\ubeta} - \ubeta^0\|= O_p(\delta)$, which completes the proof of (i).

Note that
\begin{align*}
  \|\hat{f}_k(u)- f^0_k(u)\| & = \int_{\mathcal{U}}\{\hat{f}_k(u) - f^0_k(u)\}^2du\\
  & = \int_{\mathcal{U}}\{\uB^T(u)\hat{\ugamma}_k - \uB^T(u)\ugamma^0_k + R_k(u) \}^2du\\
  & \leq 2  \int_{\mathcal{U}} \{\uB^T(u)\hat{\ugamma}_k - \uB^T(u)\ugamma^0_k\}^2du + 2  \int_{\mathcal{U}} R_k^2(u)du\\
  &=2(\hat\ugamma_k - \ugamma^0_k)^T \left(\int_{\mathcal{U}}\uB^T(u)\uB(u)du\right)(\hat\ugamma_k - \ugamma^0_k) +
  2\int_{\mathcal{U}} R_k^2(u)du.
  \end{align*}

It is obvious that $\int_{\mathcal{U}}\uB^T(u)\uB(u)du=O(1)$, so we can get
\begin{equation} \label{H1}
(\hat\ugamma_k - \ugamma^0_k)^T \left(\int_{\mathcal{U}}\uB^T(u)\uB(u)du\right)(\hat\ugamma_k - \ugamma^0_k) = O_p(n^{-2r/(2r+1)} + a^2_n).
\end{equation}

In addition, from Lemma 1, it is easy to show that
\begin{equation}\label{H2}
  \int_{\mathcal{U}} R_k^2(u)du = O_p(n^{-2r/(2r+1)}).
\end{equation}
According to (\ref{H1}) and (\ref{H2}), we complete the proof of (ii).

\textbf{Proof of Theorem 2}:
(i) Without loss of generality, we denote $\uphi = (\uphi^{nz},\uphi^{z})$, where $\uphi^{nz} = (\phi_1,\ldots, \phi_{s-1})$ and $\uphi^{z} = (\phi_s,\ldots, \phi_{q-1})$. Since $\lambda_{max}\to 0$, it can be seen $a_n = 0$ for large $n$. Then, by Theorem 1, it is sufficient to show
\[
\|\phi_j - \phi_j^0\| = O_p(n^{-r/(2r+1)}), ~~ d = 1,\ldots, s-1
\]
for $\uphi^{nz}$. For $\uphi^{z}$, for some given small $\varepsilon = Cn^{-r/(2r+1)}$, with probability approaching 1 as $n\to \infty$, for $d = s,\ldots, q - 1 $, we have
\[
\frac{\partial Q(\uphi, \ugamma)}{\partial\phi_d} > 0\ \text{when}\ 0<\phi_d<\varepsilon \text{ and } \frac{\partial Q(\uphi, \ugamma)}{\partial\phi_d} < 0\ \text{when}\ -\varepsilon<\phi_d<0.
\]

We have
$$\frac{\partial Q(\uphi, \ugamma)}{\partial\phi_d} = \frac{\partial g(\ugamma,\uphi)}{\partial\phi_d} + np_{\lambda_{3d}}(|\phi_d|)\sgn(\phi_d)$$
\begin{align*}
  \frac{\partial Q(\uphi, \ugamma)}{\partial\phi_d}
  & = \sum_{i=1}^{n}\left(Y_i - \uW_i^T(\uphi)\ugamma \right) \dot{\uW}_i^T(\uphi)\ugamma e^T_{\phi_d}\uX_i + n\dot{p}_{\lambda_{3d}}(|\phi_d|)\sgn(\phi_d) \\
  & = \sum_{i=1}^{n}\{\epsilon_i + R^T(\uX_i^T\ubeta^0)\uG_i + (I_{p+1} \otimes \uB(\uX_i\ubeta^0)\cdot \uG_i)^T(\ugamma^0 - \ugamma)\\
  & + (I_p \otimes [\uB(\uX_i^T\ubeta^0) - \uB(\uX_i^T\ubeta)]\cdot \uG_i)^T\ugamma \} \uW_i^T(\uphi)\ugamma e^T_{\phi_d}\uX_i  \\
  &+ np'_{3d}(|\phi_d|)\sgn(\phi_d)
\end{align*}
where $e_{\phi_d} = (-(1 - \|\uphi\|^2)^{-1/2}\phi_d, 0,\ldots,0,1,0,\ldots,0)^T$ with $(d+1)$th component as 1.
From conditions (\ref{A1}), (\ref{gammahatm}), (\ref{S4}) and (\ref{dotfk}), similar to \cite{Feng2013}, we have
\begin{equation}
  \frac{\partial Q(\uphi, \ugamma)}{\partial\phi_d}  = n\lambda_{3d}\{\lambda_{3d}^{-1}p'_{\lambda_{3d}}(|\phi_d|)\sgn(\phi_d) + O_p(n^{-r/(2r+1)}\lambda_{3d}^{-1})\}
\end{equation}
Clearly we can see that $\lambda_{3d}n^{r/(2r+1)} \geq \lambda_{min}n^{r/(2r+1)} \rightarrow \infty$, which implies $O_p(n^{-r/(2r+1)}\lambda_{3d}^{-1}) = o_p(1)$.
From (A6),
$\liminf_{n\to\infty}\liminf_{|\beta_{d}|\to 0^+} \frac{1}{\lambda_3}| p'_{\lambda_3}(|\beta_{d}|)| > 0$.
So we can conclude that the sign of $\frac{\partial Q(\uphi, \ugamma)}{\partial\phi_j}$ is completely determined by sign of $\phi_j$.
Hence, we prove $\hat{\beta}_j = 0$ for $j = s+1,\ldots,q$.
This completes the proof of (i).

(ii) \& (iii) Applying similar arguments as in (i), we immediately have, with probability approaching 1, $\hat{\ugamma}_{k*} = 0$ for $k = v+1,\ldots, p$ and $\hat{\gamma}_{k1} = 0 $ for $k = c+1,\ldots, p$. Then by $\sup_u\uB(u) = O(1)$ and $\hat{f}_k(\cdot) = \hat{\gamma}_{k0} + \bar{\uB}(\uX\hat{\ubeta})\hat{\ugamma}_{k*}  $, we prove $\hat{f}_k(\cdot) = c_k$ for $k = v+1,\ldots, c$ where $c_k$ is some constant and $\hat{f}_k(\cdot) = 0$ for $k = c+1,\ldots, p$.

\textbf{Proof of Theorem 3}:
Denote $\ugamma_{(\mathcal{V})} = (\ugamma_{0}^T, \ugamma_{1}^T, \ldots, \ugamma_{v}^T)^T$ and
$$\ugamma_{(\mathcal{C})} = (\ugamma_{v+1}^T, \ugamma_{v+2}^T, \ldots, \ugamma_{c}^T)^T,
\ugamma_{(\mathcal{Z})} = (\ugamma_{c+1}^T, \ugamma_{c+2}^T, \ldots, \ugamma_{p}^T)^T$$
By Theorems 1 and 2, we can see that, as $n \rightarrow \infty$, $Q(\uphi, \ugamma)$ attains the minimal value at $(\hat{\uphi}^{*T}, 0)^T$ and $(\hat{\ugamma}^{*T}_{(\mathcal{V})}, \hat{\ugamma}^{*T}_{(\mathcal{C})}, 0)^T$.
Obviously, according to (\ref{Q}), we can see that $\hat{\ugamma}^{*}_{(\mathcal{C})} = (\hat\ugamma_{v+1}^T, \hat\ugamma_{v+2}^T, \ldots, \hat\ugamma_{c+1}^T)^T$, $\hat\ugamma_k = (\hat\gamma_{k1}, 0, 0, \ldots,0)^T$ for $k = v+1, v+2, \ldots, c$.
Then, we have $\hat{f}_k(\cdot) = \hat{\gamma}_{k1}$ for $k = v+1,\ldots, c$.
Denote $\utheta^* = (\ugamma^{*T}_{(\mathcal{C})}, \uphi^{*T})^T$, and let
$$Q_{1n}(\uphi, \ugamma) = \frac{\partial Q(\uphi, \ugamma)}{\partial \ugamma^*_{(\mathcal{V})}},~~Q_{2n}(\uphi, \ugamma) = \frac{\partial Q(\uphi, \ugamma)}{\partial \ugamma^*_{(\mathcal{C})}}
, ~~ Q_{3n}(\uphi, \ugamma) = \frac{\partial Q(\uphi, \ugamma)}{\partial \uphi ^*}.~~
$$
Then, $(\hat{\uphi}^{*T}, 0)^T$ and $(\hat{\ugamma}^{*T}_{(\mathcal{V})}, \hat{\ugamma}^{*T}_{\mathcal{(C)}}, 0)^T$ must satisfy
\begin{align}\label{Q1n}
  \frac{1}{n} Q_{1n}&((\hat{\uphi}^{*T}, 0)^T, (\hat{\ugamma}^{*T}_{(\mathcal{V})}, \hat{\ugamma}^{*T}_{(\mathcal{C})}, 0)^T) \nonumber \\
  &= -\frac{2}{n} \sum_{i=1}^{n}\uW^*_{(\mathcal{V})i}(\hat{\uphi}^*) \left(Y_i - \uW^{*T}_{(\mathcal{V})i}(\hat{\uphi}^*)\hat{\ugamma}^*_{(\mathcal{V})} - \uW^{*T}_{(\mathcal{C})i} \hat{\ugamma}^*_{(\mathcal{C})}\right)
  + V_1 = 0
\end{align}
\begin{align}\label{Q2n}
  \frac{1}{n} Q_{2n}&((\hat{\uphi}^{*T}, 0)^T, (\hat{\ugamma}^{*T}_{(\mathcal{V})}, \hat{\ugamma}^{*T}_{(\mathcal{C})}, 0)^T)  \nonumber \\
   &= -\frac{2}{n} \sum_{i=1}^{n}\uW^{*}_{(\mathcal{C})i} \left(Y_i - \uW^{*T}_{(\mathcal{V})i}(\hat{\uphi}^*)\hat{\ugamma}^*_{(\mathcal{V})} - \uW^{*T}_{(\mathcal{C})i} \hat{\ugamma}^*_{(\mathcal{C})}\right) + V_2 = 0
\end{align}
\begin{align}\label{Q3n}
  \frac{1}{n} Q_{3n}&((\hat{\uphi}^{*T}, 0)^T, (\hat{\ugamma}^{*T}_{(\mathcal{V})}, \hat{\ugamma}^{*T}_{(\mathcal{C})}, 0)^T)  \nonumber \\
  &= -\frac{2}{n} \sum_{i=1}^{n}\dot{\uW}^{*T}_{(\mathcal{V})i}(\hat{\uphi}^*)\hat{\ugamma}^*_{(\mathcal{V})}J^T_{\hat{\uphi}^*}\uX^*_{(\mathcal{V})i} \left(Y_i - \uW^{*T}_{(\mathcal{V})i}(\hat{\uphi}^*)\hat{\ugamma}^*_{(\mathcal{V})} - \uW^{*T}_{(\mathcal{C})i} \hat{\ugamma}^*_{(\mathcal{C})}\right)\nonumber \\
  &+ V_3 = 0
\end{align}
where
$$V_1 = \left(0, p'_{\lambda_{11}}(\|\hat{\ugamma}_1\|)\frac{\hat{\ugamma}_1}{\|\hat{\ugamma}_1\|}, p'_{\lambda_{12}}(\|\hat{\ugamma}_2\|)\frac{\hat{\ugamma}_2}{\|\hat{\ugamma}_2\|}, \ldots, p'_{\lambda_{1v}}(\|\hat{\ugamma}_v\|)\frac{\hat{\ugamma}_v}{\|\hat{\ugamma}_v\|} \right)^T\in \mathbb{R}^{(K+h)(v+1)}$$
$$V_{2} = \left(p'_{\lambda_{2(v+1)}}(|\hat{\ugamma}_{v+1}|)\sgn(|\hat{\ugamma}_{v+1}|), \dots, p'_{\lambda_{2c}}(|\hat{\ugamma}_c|)\sgn(|\hat{\ugamma}_{c}|), 0, 0,\ldots, 0 \right)^T \in \mathbb{R}^{c-v+s-1}$$
$$V_{3} = \left(0,0,\ldots, 0, p'_{\lambda_{31}}(|\hat{\phi}_1|)\sgn(|\hat{\phi}_1|), \dots, p'_{\lambda_{3(s-1)}}(|\hat{\phi}_{s-1}|)\sgn(|\hat{\phi}_{s-1}|) \right)^T\in \mathbb{R}^{c-v+s-1}.$$

Applying Taylor expansion to $p'_{3d}(|\hat{\phi}_d|)~(d=1,\ldots,s-1)$, we get
\begin{equation}
  p'_{\lambda_{3d}}(|\hat{\phi}_d|) = p'_{\lambda_{3d}}(|\hat{\phi}_d^0|) + \{p''_{\lambda_{3d}}(|\hat{\phi}_d^0|) + o_p(1)\}(\hat{\phi}_d - \phi_d^0).
 \end{equation}
Furthermore, (A5) implies that $p''_{\lambda_{3d}}(|\hat{\phi}_d^0|) = o_p(1)$, and note that $p'_{\lambda_{3d}}(|\hat{\phi}_d^0|)=0$ as $\lambda_{\max}\rightarrow 0$.
Then, from Theorem 1 and 2, we have
$$p'_{\lambda_{3d}}(|\hat{\phi}_d|)\sgn(\hat{\phi}_d)= o_p(\hat{\phi}^* - \phi^{*})$$
Similarly, we have
$$p'_{\lambda_{1k}}(\|\hat{\ugamma}_k\|)\frac{\hat{\ugamma}_k}{\|\hat{\ugamma}_k\|}= o_p(\hat{\ugamma}^*_{(\mathcal{V})} - \ugamma^{*}_{(\mathcal{V})}),~~k=1,2,\ldots,v$$
$$p'_{\lambda_{2k}}(|\hat{\ugamma}_k|)\sgn(\hat{\ugamma}_k)= o_p(\hat{\ugamma}^*_{(\mathcal{C})} - \ugamma^{*}_{(\mathcal{C})}),~~k=v+1,\ldots,c$$


Hence, by (\ref{Q1n}) and using Taylor expansion, a simple calculation yields
\begin{align*}
  \frac{1}{n}  \sum_{i=1}^{n} & \uW^*_{(\mathcal{V})i}(\hat\uphi_0^*)  \left (Y_i - \uW^{*T}_{(\mathcal{V})i}(\hat{\uphi}^*)\hat{\ugamma}^*_{(\mathcal{V})}  - \uW^{*T}_{(\mathcal{C})i}  \hat{\ugamma}^*_{(\mathcal{C})}\right) \\
   & = \frac{1}{n}  \sum_{i=1}^{n}[\uW^*_{(\mathcal{V})i}(\uphi_0^*) + \uW^*_{(\mathcal{V})i}(\hat\uphi_0^*)
   -\uW^*_{(\mathcal{V})i}(\uphi_0^*)]\Big( \epsilon_i + \uR^T(\uphi_0^*)\uG_i^* \\
   & ~~~~- \uW_{(\mathcal{V})i}^{*T}(\uphi_0^*)(\hat{\ugamma}^*_{(\mathcal{V})} - \ugamma^*_{(\mathcal{V})}) - [\uW_{(\mathcal{V})i}^{*T}(\hat{\uphi}^*) - \uW^{*T}_{(\mathcal{V})i}(\uphi_0^*)]\hat{\ugamma}^*_{(\mathcal{V})}- \uW^{*T}_{(\mathcal{C})i} (\hat{\ugamma}^*_{(\mathcal{C})} - \hat{\ugamma}^*_{(\mathcal{C})}) \Big)\\
   & = \frac{1}{n}  \sum_{i=1}^{n}\uW^*_{(\mathcal{V})i}(\uphi_0^*)\Big( \epsilon_i + \uR^T(\uphi_0^*)\uG_i^* - \uW_{(\mathcal{V})i}^{*T}(\uphi_0^*)(\hat{\ugamma}^*_{(\mathcal{V})} - \ugamma^*_{(\mathcal{V})}) \\
   & ~~~~- [\uW_{(\mathcal{V})i}^{*T}(\hat{\uphi}^*) - \uW^{*T}_{(\mathcal{V})i}(\uphi_0^*)]\hat{\ugamma}^*_{(\mathcal{V})}- \uW^{*T}_{(\mathcal{C})i} (\hat{\ugamma}^*_{(\mathcal{C})} - \hat{\ugamma}^*_{(\mathcal{C})}) \Big) + o_p(\hat\uphi^* - \uphi_0^*)\\
   & = \frac{1}{n}  \sum_{i=1}^{n}\uW^*_{(\mathcal{V})i}(\epsilon_i + \uR^T(\uphi_0^*)\uG_i^*)-
   \frac{1}{n}  \sum_{i=1}^{n}\uW^*_{(\mathcal{V})i}(\uphi_0^*)\uW_{(\mathcal{V})i}^{*T}(\uphi_0^*)(\hat{\ugamma}^*_{(\mathcal{V})} - \ugamma^*_{(\mathcal{V})})\\
   & - \frac{1}{n}\sum_{i=1}^{n}\uW^*_{(\mathcal{V})i}[\uW_{(\mathcal{V})i}^{*T}(\hat{\uphi}^*) - \uW^{*T}_{(\mathcal{V})i}(\uphi_0^*)]\hat{\ugamma}^*_{(\mathcal{V})} - \frac{1}{n}  \sum_{i=1}^{n}\uW^*_{(\mathcal{V})i}\uW_{(\mathcal{C})i}^{*T}(\uphi_0^*)(\hat{\ugamma}^*_{(\mathcal{C})} - \ugamma^*_{(\mathcal{C})})\\
      & = \frac{1}{n}  \sum_{i=1}^{n}\uW^*_{(\mathcal{V})i}(\epsilon_i + \uR^T(\uphi_0^*)\uG_i^*)-
   \frac{1}{n}  \sum_{i=1}^{n}\uW^*_{(\mathcal{V})i}\uW_{(\mathcal{V})i}^{*T}(\uphi^*)(\hat{\ugamma}^*_{(\mathcal{V})} - \ugamma^*_{(\mathcal{V})})\\
   & - \frac{1}{n}\sum_{i=1}^{n}\uW^*_{(\mathcal{V})i}\uV^{*T}_i(\hat\uphi^* - \uphi_0^*)
   - \frac{1}{n}\sum_{i=1}^{n}
   \uW^*_{(\mathcal{V})i}\uW_{(\mathcal{C})i}^{*T}(\uphi^*)(\hat{\ugamma}^*_{(\mathcal{C})} - \ugamma^*_{(\mathcal{C})}) + o_p(\hat\uphi^* - \uphi_0^*)
\end{align*}
Then, based on (A8), Theorem 1 and $\sup_u\|\uB(u)\|=O(1)$, we have
\begin{equation}
  \hat \ugamma^*_{(\mathcal{V})} - \ugamma^*_{(\mathcal{V})} = [\uPsi_{11} + o_p(1)]^{-1}(\uLambda_{10} - \uPsi_{12}(\hat{\ugamma}^*_{(\mathcal{C})} - \hat{\ugamma}^*_{(\mathcal{C})}) - \uPsi_{13}(\hat{\uphi}^* - \uphi^*))
  \end{equation}

%
%

Thus, according to (\ref{Q2n}), we can get
\begin{align*}
  0 & =\frac{1}{n}\sum_{i=1}^{n}\uW^{*}_{(\mathcal{C})i}\left(Y_i - \uW^{*T}_{(\mathcal{V})i}(\hat{\uphi}^*)\hat{\ugamma}^*_{(\mathcal{V})} - \uW^{*T}_{(\mathcal{C})i}  \hat{\ugamma}^*_{(\mathcal{C})}\right)\\
  & = \frac{1}{n}\sum_{i=1}^{n} \uW^{*}_{(\mathcal{C})i}\Big( \epsilon_i + \uR^T(\uphi^*)\uG_i^* - \uW_{(\mathcal{V})i}^{*T}(\uphi^*)(\hat{\ugamma}^*_{(\mathcal{V})} - \ugamma^*_{(\mathcal{V})}) \\
  & ~~~~ - [\uW_{(\mathcal{V})i}^{*T}(\hat{\uphi}^*) - \uW^{*T}_{(\mathcal{V})i}(\uphi^*)]\hat{\ugamma}^*_{(\mathcal{V})}- \uW^{*T}_{(\mathcal{C})i}(\hat{\ugamma}^*_{(\mathcal{C})} - \hat{\ugamma}^*_{(\mathcal{C})}) \Big) +o_p(\hat{\ugamma}^*_{(\mathcal{C})} - \ugamma^{*}_{(\mathcal{C})})\\
  & = \frac{1}{n}\sum_{i=1}^{n} \uW^{*}_{(\mathcal{C})i}\Big( \epsilon_i + \uR^T(\uphi^*)\uG_i^* - \uW_{(\mathcal{V})i}^{*T}(\uphi^*)[\uPsi_{11} + o_p(1)]^{-1}(\uLambda_{10} - \uPsi_{12}(\hat{\ugamma}^*_{(\mathcal{C})} - \hat{\ugamma}^*_{(\mathcal{C})}) \\
  & ~~~~ - \uPsi_{13}(\hat{\uphi}^* - \uphi^*)) - [\uW_{(\mathcal{V})i}^{*T}(\hat{\uphi}^*) - \uW^{*T}_{(\mathcal{V})i}(\uphi^*)]\hat{\ugamma}^*_{(\mathcal{V})}- \uW^{*T}_{(\mathcal{C})i}(\hat{\ugamma}^*_{(\mathcal{C})} - \hat{\ugamma}^*_{(\mathcal{C})}) \Big) \\
  &~~~~ +o_p(\hat{\ugamma}^*_{(\mathcal{C})} - \ugamma^{*}_{(\mathcal{C})})\\
  & = \frac{1}{n}\sum_{i=1}^{n}\uW^{*}_{(\mathcal{C})i} \Big( \epsilon_i + \uR^T(\uphi^*)\uG_i^* - \uW_{(\mathcal{V})i}^{*T}(\uphi^*)[\uPsi_{11} + o_p(1)]^{-1}\uLambda_{10}\Big) \\
  &~~~~+ \frac{1}{n}\sum_{i=1}^{n}\uW_{(\mathcal{C})i}^{*}\uW_{(\mathcal{V})i}^{*T}(\uphi^*)[\uPsi_{11} + o_p(1)]\uPsi_{12}(\hat{\ugamma}^*_{(\mathcal{C})} - \ugamma^{*}_{(\mathcal{C})})\\
  & ~~~~ + \frac{1}{n}\sum_{i=1}^{n}\uW_{(\mathcal{C})i}^{*}\uW_{(\mathcal{V})i}^{*T}(\uphi^*)[\uPsi_{11} + o_p(1)]\uPsi_{13}(\hat{\uphi}^* - \uphi^*)\\
  & ~~~~ -  \frac{1}{n}\sum_{i=1}^{n}\uW_{(\mathcal{C})i}^{*}\Big([\uW_{(\mathcal{V})i}^{*T}(\hat{\uphi}^*) - \uW^{*T}_{(\mathcal{V})i}(\uphi^*)]\hat{\ugamma}^*_{(\mathcal{V})}  \Big) \\
  & ~~~~ - \frac{1}{n}\sum_{i=1}^{n}\uW_{(\mathcal{C})i}^{*}\uW_{(\mathcal{C})i}^{*T}(\hat{\ugamma}^*_{(\mathcal{C})} - \hat{\ugamma}^*_{(\mathcal{C})})
  + o_p(\hat{\ugamma}^*_{(\mathcal{C})} - \ugamma^{*}_{(\mathcal{C})})\\
  &\buildrel \Delta \over= J_1 + J_2 + J_3 - J_4 - J_5 +  o_p(\hat{\ugamma}^*_{(\mathcal{C})} - \ugamma^{*}_{(\mathcal{C})})
\end{align*}

Note that
\begin{equation*}
  \frac{1}{n} \sum_{i=1}^{n}\uPhi_{22}\uPhi_{11}^{-1}\uW_{(\mathcal{V})i}^{*}(\uphi^*)(\epsilon_i + \uR^T(\uphi^*)\uG_i^* - \uW_{(\mathcal{V})i}^{*T}(\uphi^*)\uPsi_{11}^{-1}\Lambda_{10})=0
\end{equation*}
\begin{equation*}
  \frac{1}{n} \sum_{i=1}^{n}(\uW_{(\mathcal{C})i}^{*}-\uPhi_{22}\uPhi_{11}^{-1}\uW_{(\mathcal{V})i}^{*}(\uphi^*))\uW_{(\mathcal{V})i}^{*T}(\uphi^*)=0
\end{equation*}
Hence, we can get
\begin{align*}
  J_1 & = \frac{1}{n} \sum_{i=1}^{n}(\uW_{(\mathcal{C})i}^{*}-\uPsi_{22}\uPsi_{11}^{-1}\uW_{(\mathcal{V})i}^{*}(\uphi^*))\epsilon_i
  \\
  & + \frac{1}{n} \sum_{i=1}^{n}(\uW_{(\mathcal{C})i}^{*}-\uPsi_{22}\uPsi_{11}^{-1}\uW_{(\mathcal{V})i}^{*}(\uphi^*))\uR(\uphi^*)\uG_i^*\\
  & + \frac{1}{n} \sum_{i=1}^{n}(\uW_{(\mathcal{C})i}^{*}-\uPsi_{22}\uPsi_{11}^{-1}\uW_{(\mathcal{V})i}^{*}(\uphi^*))
  \uW_{(\mathcal{V})i}^{*T}(\uphi^*)[\uPsi_{11}+o_p(1)]^{-1}+o_p(\hat\ugamma_{(\mathcal{C})}^* - \ugamma_{(\mathcal{C})}^*)\\
  &=\frac{1}{n} \sum_{i=1}^{n}(\uW_{(\mathcal{C})i}^{*}-\uPsi_{22}\uPsi_{11}^{-1}\uW_{(\mathcal{V})i}^{*}(\uphi^*))\epsilon_i
  + o_p(\hat\ugamma_{(\mathcal{C})}^* - \ugamma_{(\mathcal{C})}^*)
\end{align*}
Similarly, we have
\begin{align*}
  J_2 &= \uPhi_{22}\uPsi_{11}^{-1}\uPsi_{12}(\hat\ugamma_{(\mathcal{C})}^* - \ugamma_{(\mathcal{C})}^*) + o_p(\hat\ugamma_{(\mathcal{C})}^* - \ugamma_{(\mathcal{C})}^*)\\
  J_3 &= \uPsi_{22}\uPhi_{11}^{-1}\uPsi_{13}(\hat\uphi^* - \uphi^*) + o_p(\hat\uphi^* - \uphi^*)\\
  J_4 & = \uPsi_{23}(\hat\uphi^* - \uphi^*) + o_p(\hat\uphi^* - \uphi^*)\\
  J_5 &= \uPsi_{22}(\hat\ugamma_{(\mathcal{C})}^* - \ugamma_{(\mathcal{C})}^*) + o_p(\hat\ugamma_{(\mathcal{C})}^* - \ugamma_{(\mathcal{C})}^*)
\end{align*}
So we can get
\begin{align}\label{one}
  \frac{1}{n} \sum_{i=1}^{n}(\uW_{(\mathcal{C})i}^{*}&-\uPsi_{22}\uPsi_{11}^{-1}\uW_{(\mathcal{V})i}^{*}(\uphi^*))\epsilon_i \nonumber \\
  &=(\uPsi_{22}\uPsi_{11}^{-1}\uPsi_{12}-\uPsi_{22})(\hat\ugamma_{(\mathcal{C})}^* - \ugamma_{(\mathcal{C})}^*) + (\uPsi_{22}\uPsi_{11}^{-1}\uPsi_{13}-\uPsi_{23})(\hat\uphi^* - \uphi^*) \nonumber\\
  & +o_p(\hat\ugamma_{(\mathcal{C})}^* - \ugamma_{(\mathcal{C})}^*) + o_p(\hat\uphi^* - \uphi^*) \nonumber\\
  & =(\uPhi_{11}, \uPhi_{12})(\hat\utheta^*- \utheta^*) + o_p(\hat\utheta^*- \utheta^*)
\end{align}
where $\uPhi_{11} = \uPsi_{22}\uPsi_{11}^{-1}\uPsi_{12}-\uPsi_{22}, \uPhi_{12} = \uPsi_{22}\uPsi_{11}^{-1}\uPsi_{13}-\uPsi_{23}$.

According to (\ref{Q3n}), we have
\begin{align*}
  0 & =\frac{1}{n}\sum_{i=1}^{n}\hat\uV^* \left(Y_i - \uW^{*T}_{(\mathcal{V})i}(\hat{\uphi}^*)\hat{\ugamma}^*_{(\mathcal{V})} - \uW^{*T}_{(\mathcal{V})i} \hat{\ugamma}^*_{(\mathcal{C})}\right)  + o_p(\hat{\ugamma}^*_{(\mathcal{C})} - \ugamma^*_{(\mathcal{C})}) + o_p(\hat{\uphi}^*-\uphi^*)\\
  & =  \frac{1}{n}\sum_{i=1}^{n}\hat\uV^*\Big( \epsilon_i + \uR^T(\uphi^*)\uG_i^{*} - \uW_{(\mathcal{V})i}^{*T}(\uphi^*)(\hat{\ugamma}^*_{(\mathcal{V})} - \ugamma^*_{(\mathcal{V})}) - [\uW_{(\mathcal{V})i}^{*T}(\hat{\uphi}^*) - \uW^{*T}_{(\mathcal{V})i}(\uphi^*)]\hat{\ugamma}^*_{(\mathcal{V})}\\
  & ~~~~ - \uW^{*T}_{(\mathcal{C})i}(\hat{\ugamma}^*_{(\mathcal{C})} - \hat{\ugamma}^*_{(\mathcal{C})}) \Big) + o_p(\hat{\ugamma}^*_{(\mathcal{C})} - \ugamma^*_{(\mathcal{C})}) + o_p(\hat{\uphi}^*-\uphi^*)\\
  & = \frac{1}{n}\sum_{i=1}^{n}\hat\uV^* \Big(\epsilon_i + \uR^T(\uphi^*)\uG_i^* - \uW_{(\mathcal{V})i}^{*T}[\uPsi_{11} + o_p(1)]^{-1}(\Lambda_{10} - \Psi_{12}(\hat{\ugamma}^*_{(\mathcal{C})} - \hat{\ugamma}^*_{(\mathcal{C})})  \\
  &~~~~ - \Psi_{13}(\hat{\uphi}^* - \uphi^*)) -[\uW_{(\mathcal{V})i}^{*T}(\hat{\uphi}^*) - \uW^{*T}_{(\mathcal{V})i}(\uphi^*)]\hat{\ugamma}^*_{(\mathcal{V})}- \uG_i^{*T}(\hat{\ugamma}^*_{(\mathcal{C})} - \hat{\ugamma}^*_{(\mathcal{C})}) \Big)
  \\
  & ~~~~+ o_p(\hat{\ugamma}^*_{(\mathcal{C})} - \ugamma^*_{(\mathcal{C})}) + o_p(\hat{\uphi}^*-\uphi^*)\\
  & = \frac{1}{n}\sum_{i=1}^{n}  \hat\uV^*
  \Big(\epsilon_i + \uR^T(\uphi^*)\uG_i^* - \uW_{(\mathcal{V})i}^{*T}[\Psi_{11} + o_p(1)]^{-1}\Lambda_{10} \Big)  \\
  & ~~~~+ \frac{1}{n}\sum_{i=1}^{n} \hat\uV^*
  \uW_{(\mathcal{V})i}^{*T}[\Psi_{11} + o_p(1)]^{-1}\Psi_{12}(\hat{\ugamma}^*_{(\mathcal{C})} - \hat{\ugamma}^*_{(\mathcal{C})})\\
  & ~~~~+ \frac{1}{n}\sum_{i=1}^{n}  \hat\uV^*
  \uW_{(\mathcal{V})i}^{*T}[\Psi_{11} + o_p(1)]^{-1} \Psi_{13}(\hat{\uphi}^* - \uphi^*)\\
  &~~~~ -\frac{1}{n}\sum_{i=1}^{n}  \hat\uV^*
  [\uW_{(\mathcal{V})i}^{*T}(\hat{\uphi}^*) - \uW^{*T}_{(\mathcal{V})i}(\uphi^*)]\hat{\ugamma}^*_{(\mathcal{V})}  \\
  &~~~~ - \frac{1}{n}\sum_{i=1}^{n}  \hat\uV^*
 \uW^{*T}_{(\mathcal{C})i}(\hat{\ugamma}^*_{(\mathcal{C})} - \hat{\ugamma}^*_{(\mathcal{C})})
  + o_p(\hat{\ugamma}^*_{(\mathcal{C})} - \ugamma^*_{(\mathcal{C})}) + o_p(\hat{\uphi}^*-\uphi^*) \\
 &\buildrel \Delta \over=  \Delta_1 + \Delta_2 + \Delta_3 - \Delta_4 - \Delta_5  + + o_p(\hat{\ugamma}^*_{(\mathcal{C})} - \ugamma^*_{(\mathcal{C})}) + o_p(\hat{\uphi}^*-\uphi^*)
\end{align*}
where $\hat\uV^*= \dot{\uW}^{*T}_{(\mathcal{V})i}(\hat{\uphi}^*)\hat{\ugamma}^*_{(\mathcal{V})}
  J^T_{\hat{\uphi}^*}\uX^*_{(\mathcal{V})i}$.

For $\Delta_1$, we have
\begin{align*}
  \Delta_1 &= \frac{1}{n}\sum_{i=1}^{n}\hat\uV^* M_1 = \frac{1}{n}\sum_{i=1}^{n}\uV^* M_1 + \frac{1}{n}\sum_{i=1}^{n}[\dot{\uf}(\uphi^*)\uG_i^* - \dot{\uW}(\uphi^*)\ugamma^*]\uJ^T_{\hat\uphi^*}\uX_i^*M_1\\
  & + \frac{1}{n}\sum_{i=1}^{n}\dot{\uW}(\uphi^*)(\hat\ugamma^* - \ugamma^*)\uJ^T_{\hat\uphi^*}\uX_i^*M_1
  + \frac{1}{n}\sum_{i=1}^{n}[\dot{\uW}(\uphi^*) - \dot{\uW}(\hat\uphi^*)]^T\uJ^T_{\hat\uphi^*}\uX_i^*M_1\\
  &=: \Delta_{11} + \Delta_{12} + \Delta_{13} + \Delta_{14}
\end{align*}
where $M_1 = \epsilon_i + \uR^T(\uphi^*)\uG_i^* -  \uW_{(\mathcal{V})i}^{*T}(\uphi^*)\uPsi_{11}^{-1}\Lambda_{10}$.

Note that
\begin{equation*}
  \frac{1}{n} \sum_{i=1}^{n}\uPsi_{13}\uPsi_{11}^{-1}\uW_{(\mathcal{V})i}^{*}(\uphi^*)(\epsilon_i + \uR^T(\uphi^*)\uG_i^* - \uW_{(\mathcal{V})i}^{*T}(\uphi^*)\uPsi_{11}^{-1}\Lambda_{10})=0
\end{equation*}
\begin{equation*}
  \frac{1}{n} \sum_{i=1}^{n}(\uV_i^* -\uPsi_{13}^T\uPsi_{11}^{-1}\uW_{(\mathcal{V})i}^{*}(\uphi^*))\uW_{(\mathcal{V})i}^{*T}(\uphi^*)=0
\end{equation*}
Then, we can show that
\begin{align*}
  \Delta_{11} & = \frac{1}{n} \sum_{i=1}^{n}(\uV_i^*-\uPsi_{13}^T\uPsi_{11}^{-1}\uW_{(\mathcal{V})i}^{*}(\uphi^*))\epsilon_i
  \\
  & + \frac{1}{n} \sum_{i=1}^{n}(\uV_i^*-\uPsi_{13}^T\uPsi_{11}^{-1}\uW_{(\mathcal{V})i}^{*}(\uphi^*))\uR(\uphi^*)\uG_i^*\\
  & + \frac{1}{n} \sum_{i=1}^{n}(\uV_i^*-\uPsi_{13}^T\uPsi_{11}^{-1}\uW_{(\mathcal{V})i}^{*}(\uphi^*))
  \uW_{(\mathcal{V})i}^{*T}(\uphi^*)[\uPsi_{11}+o_p(1)]^{-1}+o_p(\hat\ugamma_{(\mathcal{C})}^* - \ugamma_{(\mathcal{C})}^*)\\
  &=\frac{1}{n} \sum_{i=1}^{n}(\uV_i^*-\uPsi_{13}^T\uPsi_{11}^{-1}\uW_{(\mathcal{V})i}^{*}(\uphi^*))\epsilon_i
  + o_p(\hat\ugamma_{(\mathcal{C})}^* - \ugamma_{(\mathcal{C})}^*) + o_p(\hat\uphi^* - \uphi^*)
\end{align*}
Similar to \cite{Feng2013}, we can get
$\Delta_{12} =  o_p(\hat\ugamma_{(\mathcal{C})}^* - \ugamma_{(\mathcal{C})}^*) + o_p(\hat\uphi^* - \uphi^*)$,
$$\Delta_{13} =  o_p(\hat\ugamma_{(\mathcal{C})}^* - \ugamma_{(\mathcal{C})}^*) + o_p(\hat\uphi^* - \uphi^*),
\Delta_{14} =  o_p(\hat\ugamma_{(\mathcal{C})}^* - \ugamma_{(\mathcal{C})}^*) + o_p(\hat\uphi^* - \uphi^*).$$
Hence, we have
\begin{equation}\label{Delta1}
  \Delta_1 = \frac{1}{n} \sum_{i=1}^{n}(\uV_i^*-\uPsi_{13}^T\uPsi_{11}^{-1}\uW_{(\mathcal{V})i}^{*}(\uphi^*))\epsilon_i
  + o_p(\hat\ugamma_{(\mathcal{C})}^* - \ugamma_{(\mathcal{C})}^*) + o_p(\hat\uphi^* - \uphi^*)
\end{equation}

For $\Delta_2$, we have
\begin{align*}
  \Delta_2 &= \frac{1}{n} \sum_{i=1}^{n}\hat{\uV}^*M_2
   = \frac{1}{n} \sum_{i=1}^{n}\uV^*M_2 + \frac{1}{n} \sum_{i=1}^{n} [\dot{\uf}^T(\uphi^*)\uG_i^T - \dot{\uW}(\uphi^*)\ugamma^*]\uJ^T_{\hat\uphi^*}\uX_i^* \\
  & + \frac{1}{n} \sum_{i=1}^{n}  \dot{\uW}_i^{*T}(\hat\ugamma^* - \ugamma^*)\uJ^T_{\hat\uphi^*}\uX_i^* M_2 + \frac{1}{n} \sum_{i=1}^{n} [\dot{\uW}_i^*(\uphi^*) - \dot{\uW}_i^*(\hat\uphi^*)]\uJ^T_{\hat\uphi^*}\uX_i^* M_2\\
  &\buildrel \Delta \over= \Delta_{21} +\Delta_{22}  +\Delta_{23}  +\Delta_{24}
\end{align*}
where $M_2 = \uW_{(\mathcal{V})i}^{*T}[\Psi_{11} + o_p(1)]^{-1}\Psi_{12}(\hat{\ugamma}^*_{(\mathcal{C})} - \hat{\ugamma}^*_{(\mathcal{C})})$.
Hence, we have
\begin{equation*}
  \Delta_{21} =  \uPsi_{13}^T \Psi_{11}^{-1}\Psi_{12}(\hat{\ugamma}^*_{(\mathcal{C})} - \hat{\ugamma}^*_{(\mathcal{C})})  + o_p(\hat{\ugamma}^*_{(\mathcal{C})} - \hat{\ugamma}^*_{(\mathcal{C})})
\end{equation*}
Similar arguments to that of $J_{12}$, we have
\begin{equation*}
  \Delta_{22} =  o_p(\hat{\ugamma}^*_{(\mathcal{C})} - \hat{\ugamma}^*_{(\mathcal{C})}) ,
    \Delta_{23} =  o_p(\hat{\ugamma}^*_{(\mathcal{C})} - \hat{\ugamma}^*_{(\mathcal{C})}), ~{\rm and}~
  \Delta_{24} =  o_p(\hat{\ugamma}^*_{(\mathcal{C})} - \hat{\ugamma}^*_{(\mathcal{C})}) .
\end{equation*}
Therefore, we have
\begin{equation}\label{Delta2}
  \Delta_{2} = \uPsi_{13}^T \uPsi_{11}^{-1}\uPsi_{12}(\hat{\ugamma}^*_{(\mathcal{C})} - \hat{\ugamma}^*_{(\mathcal{C})})  + o_p(\hat{\ugamma}^*_{(\mathcal{C})} - \hat{\ugamma}^*_{(\mathcal{C})}).
\end{equation}
Similarly, we have
\begin{equation}\label{Delta3}
  \Delta_{3} = \uPsi_{13}^T \uPsi_{11}^{-1}\uPsi_{13}(\hat\uphi^* - \uphi^*)  + o_p(\hat\uphi^* - \uphi^*).
\end{equation}

Now we consider $\Delta_4$, applying Taylor expansion, we have
\begin{align*}
  \Delta_4 & = \frac{1}{n}\sum_{i=1}^{n}  \hat\uV^*
  [\uW_{(\mathcal{V})i}^{*T}(\hat{\uphi}^*) - \uW^{*T}_{(\mathcal{V})i}(\uphi^*)]\hat{\ugamma}^*_{(\mathcal{V})}\\
  & = \frac{1}{n}\sum_{i=1}^{n}\hat\uV^*[\dot{\uW}_i^{*T}(\uphi^*)\hat\ugamma_{(\mathcal{V})i}\uJ^T_{\uphi^*}\uX_i^{*T}
  (\hat\uphi^* - \uphi) + o_p(\hat\uphi^* - \uphi)]\\
  & = \frac{1}{n}\sum_{i=1}^{n}\hat\uV^*[\uV^{*T}(\hat\uphi^* - \uphi^*)) + o_p(\hat\uphi^* - \uphi^*)]\\
  & = \frac{1}{n}\sum_{i=1}^{n}\uV^*\uV^{*T}(\hat\uphi^* - \uphi^*) + \frac{1}{n}\sum_{i=1}^{n}\uV^*(\hat\uphi^* - \uphi^*)\dot{\uW}_i^{*T}(\uphi^*)(\hat\ugamma^*_{(\mathcal{V})}- \ugamma^*_{(\mathcal{V})})\uJ^T_{\hat{\uphi}^*}\uX_i^*\\
  & + \frac{1}{n}\sum_{i=1}^{n}\uV^{*T}(\hat\uphi^* - \uphi^*)[\dot{\uf}(\uphi^*)\uG_i^* - \dot{\uW}_i^{*T}(\uphi^*)\ugamma^*_{(\mathcal{V})}]\uJ^T_{\hat{\uphi}^*}\uX_i^*\\
  &=\frac{1}{n}\sum_{i=1}^{n}\uV^*\uV^{*T}(\hat\uphi^* - \uphi^*) + o_p(\hat\uphi^* - \uphi^*)\\
  &=\uPsi_{21}(\hat\uphi^* - \uphi^*) + o_p(\hat\uphi^* - \uphi^*).
\end{align*}

Similarly, we have
\begin{equation}\label{Delta5}
  \Delta_5 = \uPsi_{23}^T(\hat\ugamma^*_{(\mathcal{C})} - \ugamma^*_{(\mathcal{C})}) + o_p(\hat\ugamma^*_{(\mathcal{C})} - \ugamma^*_{(\mathcal{C})})
\end{equation}

So we can get
\begin{align}\label{two}
  \frac{1}{n}& \sum_{i=1}^{n}(\uV_i^*  -\uPsi_{13}^T\uPsi_{11}^{-1}\uW_{(\mathcal{V})i}^{*}(\uphi^*))\epsilon_i\\
   &= \uPsi_{23}^T(\hat\ugamma^*_{(\mathcal{C})} - \ugamma^*_{(\mathcal{C})}) + \uPsi_{21}^T(\hat\uphi^* -
   \uphi^*) -\uPsi_{13}^T \uPsi_{11}^{-1}\uPsi_{12}(\hat{\ugamma}^*_{(\mathcal{C})} - \hat{\ugamma}^*_{(\mathcal{C})}) \nonumber\\
  & -\uPsi_{13}^T \uPsi_{11}^{-1}\uPsi_{13}(\hat\uphi^* - \uphi^*)
  + o_p(\hat\ugamma^*_{(\mathcal{C})} - \ugamma^*_{(\mathcal{C})})+ o_p(\hat\uphi^* - \uphi^*) \nonumber\\
  & = (\uPsi_{21}^T - \uPsi_{13}^T\uPsi_{11}\uPsi_{13})(\hat\uphi^* - \uphi^*) + (\uPsi_{23}^T - \uPsi_{13}^T\uPsi_{11}\uPsi_{12})(\hat\ugamma^*_{(\mathcal{C})} - \ugamma^*_{(\mathcal{C})}) \nonumber \\
  & + o_p(\hat\ugamma^*_{(\mathcal{C})} - \ugamma^*_{(\mathcal{C})})+ o_p(\hat\uphi^* - \uphi^*) \nonumber \\
  & = (\uPhi_{21}, \uPhi_{22})(\hat\utheta^*-\utheta^*) + o_p(\hat\utheta^*-\utheta^*).
\end{align}
where $\uPhi_{21} = \uPsi_{23}^T - \uPsi_{13}^T\uPsi_{11}\uPsi_{12}$ and $\uPhi_{22} = \uPsi_{21}^T - \uPsi_{13}^T\uPsi_{11}\uPsi_{13}$.

According to (\ref{one}) and (\ref{two}), we have
\begin{align}\label{three}
  \sqrt n (\hat \utheta^*  - \utheta ) &= {\left( {\begin{array}{*{20}{c}}
{{\uPhi _{11}}}&{{\uPhi _{12}}}\\
{{\uPhi _{21}}}&{{\uPhi _{22}}}
\end{array}} \right)^{-1}}\frac{1}{{\sqrt n }}\sum\limits_{i = 1}^n {\left( {\begin{array}{*{20}{c}}
\uW_{(\mathcal{C})i}^{*}-\uPsi_{22}\uPsi_{11}^{-1}\uW_{(\mathcal{V})i}^{*}(\uphi^*)\\
\uV_i^*  -\uPsi_{13}^T\uPsi_{11}^{-1}\uW_{(\mathcal{V})i}^{*}(\uphi^*)
\end{array}} \right){\epsilon_i}} \nonumber \\
&+ o_p(1).
\end{align}
By the central limit theorem and Slutsky's theorem, we can see that $\hat \utheta^*$ is consistent and has asymptotic normality.

It follows from (\ref{JacobM}) that
$$\hat\ubeta^* - \ubeta^* = \uJ_{\uphi^*}(\hat\uphi^* - \uphi^*) + O_p(n^{-1}).$$
Hence, we can get
$$\sqrt{n}(\hat\uvartheta^* - \uvartheta^*) = \left( {\begin{array}{*{20}{c}}
1&0\\
0&\uJ_{\uphi^*}
\end{array}} \right)\sqrt n (\hat \utheta  - \utheta ).$$
Therefore, we can get the asymptotic covariance matrix $\uSigma$ as
$$\uSigma = \left( {\begin{array}{*{20}{c}}
\Sigma_1^{-1}&\uzero\\
\uzero& \uJ_{\uphi_0^*}\Sigma_2^{-1}\uJ^T_{\uphi_0^*}
\end{array}} \right)$$
Then, the proof of theorem 3 is completed.

\section*{Acknowledgments}
This work was supported in part by the National Institutes of Health [R21HG010073 to Y.C.], the University Social Science Research Project of Anhui Province (SK2020A0051 to M.Z.) and the Social Science Foundation of Ministry of Education of China [21YJAZH081 and 19YJCZH250 to M.Z.]. The funding agencies had no role in study design and data collection, analysis and interpretation.

\end{document}